\documentclass[12pt]{article}

\usepackage{amsmath,amssymb,amscd}
\usepackage{graphicx}

\makeatletter
 
  \@addtoreset{equation}{section}
 \makeatother

\newcommand{\be}{\begin{equation}}
\newcommand{\ee}{\end{equation}}
\newcommand{\bea}{\begin{eqnarray}}
\newcommand{\eea}{\end{eqnarray}}
\newcommand{\beann}{\begin{eqnarray*}}
\newcommand{\eeann}{\end{eqnarray*}}
\newcommand{\nn}{\nonumber}
\newcommand{\ba}{\begin{array}}
\newcommand{\ea}{\end{array}}

\DeclareMathOperator{\Tr}{Tr}
\DeclareMathOperator{\Det}{Det}

\DeclareMathOperator{\rank}{rank}

\newcommand{\e}{\epsilon}

\newcommand{\Z}{\mathbb{Z}}

\newcommand{\C}{\mathbb{C}}

\newcommand{\del}{\partial}
\newcommand{\D}{{\cal D}}

\newcommand{\B}{{\cal B}}
\newcommand{\F}{{\cal F}}

\newcommand{\Op}{{\cal O}}
\newcommand{\cL}{{\cal L}}
\newcommand{\Phib}{\bar{\Phi}}
\newcommand{\Phit}{\tilde{\Phi}}
\newcommand{\Xt}{\tilde{X}}
\newcommand{\psit}{\tilde{\psi}}

\newcommand{\calZ}{{\cal Z}}

\newcommand{\Kahler}{K\"ahler }

\newcommand{\zb}{{\bar{z}}}

\newcommand{\eb}{{\bar{\e}}}

\newcommand{\Vol}{{\rm Vol}}
\newcommand{\A}{\mathcal{A}}

\newcommand{\N}{{\cal N}}
\newcommand{\calP}{{\cal P}}

\newcommand{\kdk}{\kappa\wedge d\kappa}

\def\XXint#1#2#3{{\setbox0=\hbox{$#1{#2#3}{\int}$} 
\vcenter{\hbox{$#2#3$}}\kern-.5\wd0}}

\begin{document}

\setlength{\oddsidemargin}{0cm}
\setlength{\baselineskip}{7mm}

\begin{titlepage}
\renewcommand{\thefootnote}{\fnsymbol{footnote}}
\begin{normalsize}
\begin{flushright}
\begin{tabular}{l}
KEK-TH-1542\\
May, 2012
\end{tabular}
\end{flushright}
  \end{normalsize}

~~\\

\vspace*{0cm}
    \begin{Large}
    \begin{bf}
       \begin{center}
         {Non-Abelian Localization for Supersymmetric Yang-Mills-Chern-Simons Theories on Seifert Manifold}
       \end{center}
    \end{bf}   
    \end{Large}
\vspace{1cm}

\begin{center}
Kazutoshi O{\sc hta}$^{1)}$\footnote
            {
e-mail address : 
kohta@law.meijigakuin.ac.jp}
    {\sc and}
Yutaka Y{\sc oshida}$^{2)}$\footnote
            {
e-mail address : 
yyoshida@post.kek.jp
}

\vspace{0.7cm}

       $^{1)}$ {\it Institute of Physics, Meiji Gakuin University, 
Yokohama 244-8539, Japan}\\
      \vspace{0.3cm}
       $^{2)}$ {\it High Energy Accelerator Research Organization (KEK), \\
Tsukuba, Ibaraki 305-0801, Japan }

\end{center}

\vspace{1cm}

\begin{abstract}
We derive non-Abelian localization formulae for
supersymmetric Yang-Mills-Chern-Simons theory with matters on a Seifert manifold $M$,
which is the three-dimensional space of a circle bundle over a two-dimensional Riemann surface $\Sigma$,
by using the cohomological approach introduced by K\"all\'en.
We find that the partition function and the vev of the supersymmetric Wilson loop
reduces to a finite dimensional integral and
summation over classical flux configurations labeled by discrete integers.
We also find the partition function reduces further to just a discrete sum over integers
in some cases, and evaluate the supersymmetric index (Witten index) exactly
on $S^1\times \Sigma$.
The index completely agrees with the previous prediction from field theory
and branes. We discuss a vacuum structure of the ABJM theory
deduced from the localization.

\end{abstract}

\vfill
\end{titlepage}
\vfil\eject

\setcounter{footnote}{0}

\section{Introduction}

The localization theorem (Duistermaat-Heckman formula) \cite{Duistermaat:1982}
(see also \cite{Karki:1993bw}) reduces an integral over a symplectic
manifold to a piecewise integral over isometry fixed locus.
If the fixed locus is a set of isolated points, the integral reduces further to a discrete sum.
This integral formula is so powerful and applied to various problems in mathematics and physics.
In particular, in the last decade, the localization theorem has been utilized in physics for
the evaluation of the volume of the symplectic moduli space \cite{Moore:1997dj,Gerasimov:2007ap,Miyake:2011yr}, or
the non-perturbative corrections to the supersymmetric gauge theories \cite{Nekrasov:2003rj}.
More recently, the localization theorem is applied to the exact calculation of the partition function or
vev of the Wilson loop in the supersymmetric gauge theories on a compact sphere \cite{Pestun:2007rz,Kapustin:2009kz}
and leads the significant developments in both gauge and string theory.

The {\it equivariant cohomology} plays a key role in the proof of the localization theorem \cite{Atiyah:1984}.
The equivariant cohomology is constructed from the operator
\be
d_V = d + \iota_V,
\ee
which is the sum of the exterior derivative $d$ and the interior product $\iota_V$ associated
with the (Hamiltonian) vector field $V$.
$d_V$ does not make a cohomology in a usual sense, since
$d_V$ is not nilpotent
\be
d_V^2 = \cL_V,
\ee 
where $\cL_V$ is the Lie derivative along the vector field $V$.
However, if we restrict the differential forms, on which $d_V$ is acting,
to the invariant ones under the Lie derivative $\cL_V$,
then the $d_V$ becomes nilpotent and make a cohomology,
which is called the equivariant cohomology.
The localization theorem says that the integral on the symplectic manifold,
which has the isometry generated by $\cL_V$,
reduces on the fixed locus of the vector field.

It is known that the physical theories equipped with the supersymmetry 
is closely related to mathematical theory of the equivariant cohomology.
Indeed, the square of the supercharge generates the sum of the possible symmetries in
the supersymmetric theory, like the Poincar\'e (with dilatation if theory is superconformal),
gauge and R-symmetry transformations.
In this sense, we can identify the supercharges with the above $d_V$.
It is also natural in physics to restrict the observables on the symmetry invariant ones,
in particular the gauge invariant operators,
as well as the construction of the equivariant cohomology.
It is difficult in general to use all of symmetries for the equivariant cohomology and localization.
So we have to reduce some of symmetries by choosing specific supercharges.
This procedure is called the topological twisting and the constructed theory is called
the topological field theory or cohomological field theory in the sense of the equivariant cohomology.
The cohomological field theory remarkably develops itself and has brought about
many benefits in physics and mathematics like the calculation of the topological invariants.
 
The other restriction of the supercharges occurs on the curved manifold.
The (rigid) supersymmetry breaks in general on the curved manifold since
the (global) Poincar\'e symmetry is broken.
However, if there exist some isometries on the curved space,
some supercharges, which is the Killing spinors associated with the isometry, may still remain.
Then we can construct the equivariant cohomology by using at least one residual supercharge.
If the square of the residual supercharge generates the isometry translation (Lie derivative)
and gauge transformation,
we can discuss the localization in the supersymmetric gauge theory for the isometry
and gauge invariant states by using the equivariant cohomology.
We can see clearly the meanings of the isometry fixed points and
 the role of the equivariant cohomology in the localization of the supersymmetric gauge theory.

On the other hand, in the discussion of the localization in three-dimensional supersymmetric Chern-Simons (CS)
gauge theory with matters \cite{Kapustin:2009kz,Hama:2010av} (see for review \cite{Marino:2011nm}),
the supercharge used for the localization is chosen to be nilpotent itself ($Q^2=0$)
while the supercharge in the cohomological field theory generates the isometry translation.
Of course, the arguments of the localization by the nilpotent supercharge
such as the 1-loop (WKB) exactness of the partition function is precise,
and the results on the partition function and Wilson loops are exact,
since their quantities are independent of the choice of the supercharge.
The localization fixed locus can be also read from the supersymmetric transformations.
There is nothing to lose in the nilpotent supercharge formulation
of the localization in the supersymmetric gauge
theory on the specific manifold like $S^3$.
However the formulation in \cite{Kapustin:2009kz} seems to relay on the specific metric on $S^3$
although the CS theory is essentially topological.
For other kinds of manifold which have the same topology but different metric as $S^3$,
the other discussions on the determinants of the harmonics are needed
\cite{Gang:2009wy,Hama:2011ea,Imamura:2011wg}.
The nilpotent supercharge does not also explain why the supersymmetric Wilson loop along
the isometry of the space is rather special in the localization.

The essential symmetry of the localization in the supersymmetric CS theory 
is the $U(1)$ isometry of the Hopf fibration of $S^3$.
The Wilson line along the $U(1)$ isometry is special since it preserves the isometry.
K\"all\'en recently has pointed out there is alternative choice of the supercharge
which respects to the isometry of the manifold,
and reformulate the localization in the supersymmetric CS theory on more general three-dimensional
Seifert manifold $M$,
which is a non-trivial circle bundle over a Riemann surface $\Sigma$
and possesses at least the $U(1)$ isometry \cite{Kallen:2011ny}.
In his formulation, the supercharge is not nilpotent but the square of the supercharge generates
the Lie derivative along the isometry and the gauge transformation.
In this sense, this formulation is very suitable for considering the relation to
the equivariant cohomology.
Indeed,
the 1-loop determinants of the fields can be determined only by the topological nature (index theorem)
in this formulation.
This cohomological formulation also explains the exact partition function for the
bosonic (non-supersymmetric) CS theory on the Seifert manifold
\cite{Witten:1988hf,Rozansky:1994wv,Lawrence:2004,Marino:2002fk,Aganagic:2002wv}
(see also for good reviews \cite{Marino:2004uf,Marino:2004eq}),
as a consequence of the non-Abelian localization
\cite{Beasley:2005vf,Blau:2006gh}.

In this paper, we follow K\"all\'en's cohomological formulation of the localization
in the supersymmetric CS theory on the Seifert manifold $M$
and generalize to include the matter multiplets giving in the ABJM
theory \cite{Aharony:2008ug,Aharony:2008gk}.
We construct the cohomological theory from
the supersymmetry on $M$ by choosing a specific supercharge $Q$
(BRST charge)
respecting the isometry.
We more generally would like to consider Yang-Mills-Chern-Simons (YMCS) theory as displayed in the title,
since dynamics of YMCS theory is perfectly interpreted by a brane configuration with $(p,q)$5-branes
in Type IIB string theory
\cite{Kitao:1998mf,Ohta:1999gj,Bergman:1999na,Ohta:1999iv}.
YMCS theory contains the Yang-Mills (YM) action, which is
the kinetic terms of the gauge fields, in addition to the topological CS term.
However, as we will see, the YM action is written as an exact form of the supercharge $Q$.
The cohomological formulation says that the $Q$-exact action does not change the
partition function or vev of the cohomological observables,
that is, the partition function or vev of YMCS theory is essentially equivalent to
ones in  theory with the CS terms alone.
In this sense, we will just treat CS theory without the YM action, but
the YM action plays the essential role when one evaluates the 1-loop determinants
and derive the localization formula.
So it makes sense to consider the YMCS theory though we are interested in
the low energy (superconformal) CS theory,
in the context of the cohomological field theory and localization arguments.

The paper is organized as follows:

In section 2, we briefly review some basics and formulae on the Seifert manifold.
In this paper, we concentrate on a special Seifert manifold which is the circle bundle
over the smooth Riemann surface (non-orbifolding case).
So we explain the properties of the Seifert manifold for this restricted case.
We succeedingly construct the BRST transformations from the supercharges,
which is compatible with the $U(1)$ isometry on the Seifert manifold.
We also discuss the cohomological properties of the supersymmetric CS term and Wilson loop.
They play important roles in the following discussions.

In section 3, we derive the localization formula for the vector multiplet
by using the BRST transformations and cohomological observables.
We can determines a general formula for the 1-loop determinants
which dominates in the localization.
In the cohomological field theory approach, the evaluation of the 1-loop
determinants is simpler rather than the counting the spherical harmonics,
since the determinants are essentially infinite products modes on the two-dimensional
Riemann surface and Kaluza-Klein (KK) modes on the $S^1$ fiber. 
That is, CS theory on $M$ is equivalent to the two-dimensional (cohomological) YM
theory \cite{Witten:1992xu,Blau:1995rs} with the infinite number of the KK fields,
and each number of zero modes can be determined by
the index theorem (Hirzebruch-Riemann-Roch theorem) on $\Sigma$.
Thus we can exactly evaluate the 1-loop determinants and localization formula for
the supersymmetric CS theory on $M$.
We find that there is no quantum level shift of the CS coupling in the supersymmetric
CS theory without matter,
in contrast with the bosonic CS theory,
since the vector multiplet is ruled by the D-term which is
a {\it real} object and the determinants do not admit phases.
The exact partition function obtained by the localization also tells us
various interesting informations on the vacuum structure of the CS theory.
In particular, if we consider the partition function of the supersymmetric CS theory
on $S^1\times \Sigma$, which is a special case of the Seifert manifold,
we can evaluate the supersymmetric index (Witten index) of the system.
It coincides with the known results from the field theory \cite{Witten:1999ds}
 and branes in string theory \cite{Ohta:1999iv}.
We also discuss the partition functions in the various limit of the degree of the fiber
and the genus of $\Sigma$.

In section 4, we introduce matter chiral superfields in various representations.
The matter superfields can get explicit mass or anomalous dimension for the superconformal theory.
We find that the mass or anomalous dimension
is embodied as the twisted boundary condition along the $S^1$ fiber.
Then we can treat the matter superfields in the framework of the equivariant cohomology
as well as the vector multiplet.
The chiral superfields are ruled by {\it complex} F-terms.
So the 1-loop determinants for the chiral superfields
admit phases
and may shift quantum mechanically the CS level.
We can show the phases are cancelled out in the self-conjugate representations
similar to the case on $S^3$.
Finally we give the exact partition function for the ABJM theory (and its generalization)
on the Seifert manifold
and discuss the properties and the vacuum structures of it.

The last section is devoted to conclusion and discussion.

\section{Cohomological Field Theory on Seifert Manifold}

\subsection{Basics of the Seifert manifold}

We consider the supersymmetric YMCS theories on a three-dimensional
manifold $M$, which is compact and orientable.
The three-manifold $M$ admits a contact structure, which means that
a three-form $\kdk$ does not vanish anywhere on $M$.
The 1-form $\kappa$ is called a contact form.
The three-form $\kdk$ plays an analogous role to a symplectic form on
even dimensional manifold.

We also require that $M$ admits a free $U(1)$ action, that is,
$M$ has a $U(1)$ isometry. In this case, $M$ can be represented by
the total space of a circle ($S^1$) bundle over a Riemann surface $\Sigma$,
\be
\begin{CD}
S^1 @>p>> M \\
@. @VV{\pi}V\\
 @. \Sigma
\end{CD}
\qquad ,
\ee
where we have determined that the degree of the bundle is $p$ and
the genus (the number of handles) of the Riemann surface is $h$.
The free $U(1)$ action obviously acts on the circle fiber.
This kind of three-manifold is called a Seifert manifold.

Precisely speaking, Seifert manifolds are defined by
more general principal $U(1)$-bundles over a Riemann surface with an orbifold,
where there are some fixed points of the $U(1)$ action.
We can extend the following results to the orbifold cases,
but for simplicity of explanations and because of interests as physical models,
we will take into account the non-orbifold case only in the following discussions.

The contact form $\kappa$ is associated with the circle fiber direction.
Since $M$ has the $U(1)$ isometry, there exists a vector field $V$
along the circle direction. Then $\kappa$ and $V$ satisfy
\be
\iota_V \kappa = 1,
\label{iVk=1}
\ee
where $\iota_V$ stands for the interior product with $V$.
This means that the 1-form $\kappa$ is a base of $S^1$.
We normalize $\kappa$ to be
\be
\int_{S^1} \kappa = 2\pi \ell
\ee
where $\ell$ is the radius of the fiber $S^1$.

On the other hand,
the base for a residual part of $M$ transverse to $\kappa$ contains in $d\kappa$,
namely
\be
d\kappa = \frac{p}{2\ell} \pi^* \omega,
\ee
where $p$ is the degree of the fiber bundle and $\pi^*\omega$ is the pullback of 
the symplectic (volume) form on the base Riemann surface $\Sigma$.
The transversality between $\kappa$ and $d\kappa$ says $\iota_V d\kappa = 0$,
namely combining with (\ref{iVk=1}), we find
\be
\cL_V \kappa = 0,
\ee
where $\cL_V\equiv d \iota_V+\iota_V d$ is the Lie derivative along $V$.
Using the fact that the symplectic form on $\Sigma$ gives the area $\A$ of $\Sigma$
\be
\int_{\Sigma} \omega = \A,
\ee
the integral of $\kdk$ over $M$ reduces to
\be
\int_M \kappa \wedge d\kappa
=\frac{p}{2\ell}\int_{S^1} \kappa \int_\Sigma\omega =p \pi \A.
\ee
In this sense, $\frac{\ell}{2} \kdk$ can be regarded as the volume form $\star 1$ on $M$.
Here $\star$ is the Hodge star operator defined on $M$.

The metric also can be decomposed into a diagonal form
\be
ds_M^2 = \pi^*ds_{\Sigma}^2 + \kappa \otimes \kappa,
\ee
where $\pi^*ds_{\Sigma}^2$ is the pullback of a conformally flat metric on $\Sigma$
with complex coordinates $(z,\zb)$
\be
ds_{\Sigma}^2 = \Omega dz \otimes d\zb.
\label{metric}
\ee
So we can almost treat $M$ as the direct product of the Riemann
surface $\Sigma$ and the $S^1$ fiber of the $\kappa$-direction.
(Of course, since $\kappa$ is a function of points on $\Sigma$,
$M$ is not the direct product.)
This fact will be useful to construct supersymmetry on $M$.

Using the above properties, any 1-form (vector field) $A$ on $M$
 can be decompose into
\be
A = A_\Sigma +  A_\kappa \kappa,
\ee
where $A_\kappa=\iota_V A$ is a scalar component along $\kappa$
and $A_\Sigma$ is a 1-form component horizontal to it.
By definition, we find $\iota_V A_\Sigma=0$,
then $\iota_V d A_\Sigma = \cL_V A_\Sigma$.
More generally, any $n$-form $\Omega$ on $M$ is decomposed by
using a projector $P\equiv \kappa \wedge \iota_V$
\be
\Omega = (1-P)\Omega + P\Omega.
\ee
Thus $\iota_V \Omega$ is a $(n-1)$-form component along
the fiber direction.
Similarly, the exterior derivative can be decomposed into
\be
d = (1-P)d + P d = \pi^* d_\Sigma + \kappa \wedge \iota_V d,
\ee
where $\pi^* d_\Sigma$ is the pullback of the exterior derivative on $\Sigma$.
Using these definitions, one can see
\bea
A_\Sigma \wedge d A_\Sigma
&=&A_\Sigma \wedge (\pi^* d_\Sigma A_\Sigma + \kappa \wedge \iota_V d A_\Sigma)\nn\\
&=&A_\Sigma \wedge \kappa \wedge \cL_V A_\Sigma.
\eea
We will use these properties in the following sections.

\subsection{Supersymmetry and equivariant cohomology}

Let us now construct supersymmetric gauge theories
on the Seifert manifold $M$.
Three-dimensional $\N=2$ supersymmetric gauge theory,
which has four supercharges and is obtained from a
dimensional reduction from four-dimensional $\N=1$ supersymmetric
gauge theory with the gauge group $G$, contains
a vector field $A_\mu$ ($\mu=1,2,3$),
two-component Majorara fermions (gauginos) $\lambda,\tilde{\lambda}$,
a hermite scalar field in the adjoint representation $\sigma$,
and an auxiliary field $D$. This field multiplet is called a vector multiplet. 
We can also include matter chiral superfields which are coupled with
the vector multiplets, but we concentrate only on the theory
of the vector multiplets (pure gauge theory) for a while.
Three-dimensional supersymmetric gauge theories including matter multiplets
will be discussed in the next section.

The supersymmetric YM
action which is invariant under the supersymmetry 
on the curved manifold $M$ with a Euclidian signature metric
is written in terms of the fields in the vector multiplets
\begin{multline}
S_\text{SYM} = \frac{1}{g^2}\int_M d^3x \sqrt{g}
\Tr\Bigg[
\frac{1}{2}F_{\mu\nu}F^{\mu\nu}
+\D_\mu\sigma D^\mu \sigma
+\left(D+\frac{1}{\ell}\sigma\right)^2\\
+i\bar{\lambda}\gamma^\mu\D_\mu\lambda
+i\bar{\lambda}[\sigma,\lambda]-\frac{1}{2\ell}\tilde{\lambda}{\lambda}
\Bigg]
.
\label{SYM action}
\end{multline}
In three-dimensions, we can also include the supersymmetric CS term
in the action
\be
S_\text{SCS} = \frac{k}{4\pi}\int_M
\Tr\Big[
A\wedge dA -\frac{2i}{3}A\wedge A \wedge A
\Big]
+\frac{k}{4\pi}\int_M d^3x \sqrt{g}
\Tr\Big[
2 D \sigma
-\tilde{\lambda}\lambda
\Big]
,
\label{SCS action}
\ee
which is invariant under the supersymmetric transformations.
If we take into account the both actions $S_\text{SYM}$ and $S_\text{SCS}$
at the same time, the system is called supersymmetric YMCS theory.
We discuss the exact partition function and
dynamics of this system via the localization.


In general, the supersymmetry  is violated due to
the curvature of $M$.
However, at least, one of the supersymmetries survives since $M$
has the $U(1)$ isometry after twisting the fields topologically.
So if we choose suitably one of the supercharges,
which respects the isometry, one can obtain the residual symmetry
of the system.


Redefining the fields up to
irrelevant overall constants and phases,
we obtain the following transformations from the
supersymmetry (see Appendix A) 
\be
\begin{array}{ll}
Q A_z = \lambda_z, & Q \lambda_z =-\D_z\sigma+iF_{z \kappa},\\
Q A_\zb =  \lambda_\zb,& Q \lambda_\zb = -\D_\zb \sigma+i F_{\zb \kappa},\\ 
Q A_\kappa = \eta,\\
Q \sigma = i \eta, & Q \eta =  -\D_\kappa \sigma\\
Q D =2i(\D_z \lambda_\zb-\D_\zb \lambda_z)
 -i\D_\kappa\chi_r-i[\sigma,\chi_r]-\frac{i}{\ell}\eta,
 & Q \chi_r = -2iF_{z\zb}+D+\frac{1}{\ell}\sigma.
\label{BRST transformations 1}
\end{array}
\ee
This kind of the transformations is called the BRST transformations
conventionally in topological field theory.
Introducing a form notation on $M$,
which is
$A=A_zdz + A_\zb d\zb +A_\kappa \kappa$ and
$\lambda=\lambda_zdz + \lambda_\zb d\zb + \eta \kappa$,
we can rewrite the BRST transformations
(\ref{BRST transformations 1}) compactly to
\be
\begin{array}{ll}
Q A = \lambda, & Q \lambda = -d_A \sigma -i\iota_V F,\\
Q \sigma = i \eta, \\
Q D = \frac{2i}{\ell}\frac{\kappa\wedge d_A\lambda}{\kappa\wedge d\kappa}
-i( \iota_V  d_A \chi_r +[\sigma,\chi_r])
-\frac{i}{\ell}\eta,
& Q \chi_r = -\frac{2i}{\ell}\frac{\kappa\wedge F}{\kappa\wedge d\kappa}
+D+\frac{1}{\ell}\sigma,
\end{array}
\ee
where $d_A \sigma = d\sigma - i [A,\sigma]$ and $F = dA - i A \wedge A$
and  $\frac{\kappa\wedge F}{\kappa\wedge d\kappa}$ is the formal notation
used in \cite{Beasley:2005vf} with the normalization of the volume form $\star 1 = \frac{\ell}{2}\kdk$.

Furthermore, if we define a shift of the D-field by
$Y_r \equiv -\frac{2i}{\ell}\frac{\kappa\wedge F}{\kappa\wedge d\kappa}
+D+\frac{1}{\ell}\sigma$,
then we obtain
\be
\begin{array}{ll}
Q A = \lambda, & Q \lambda = -i( \iota_V F -id_A \sigma),\\
Q \sigma = i \eta, \\
Q Y_r =-i(\iota_V d_A \chi_r +[\sigma,\chi_r]), & Q \chi_r =Y_r.
\label{BRST transformations form}
\end{array}
\ee
This simple transformation law becomes more interesting
by introducing a combination of scalar fields in the adjoint representation
\be
\Phi \equiv A_\kappa +  i\sigma.
\label{definition of Phi}
\ee
Noting that $\eta = \iota_V \lambda$ and
$Q A_\kappa = \iota_V (QA) = \eta$, we find immediately
\be
Q \Phi=0.
\ee
Using these notations, the BRST transformations are finally reduced to
\be
\begin{array}{ll}
Q A = \lambda, & Q \lambda = -i(\cL_V- \delta_\Phi )A ,\\
Q \Phi = 0, \\
Q Y_r = -i(\cL_V-\delta_\Phi)\chi_r, & Q \chi_r =Y_r,
\end{array}
\label{equivariant cohomology}
\ee
where
we have defined a gauge transformation by regarding formally $\Phi$ as a gauge parameter,
that is,
\bea
&& \delta_\Phi A = d_A \Phi = d_A A_\kappa +i d_A \sigma,\\
&& \delta_\Phi \chi_r = i[\Phi,\chi_r] = i[A_\kappa,\chi_r]-[\sigma,\chi_r].
\eea
Thus we find that the square of the transformations by $Q$ generates
\be
Q^2 = -i(\cL_V -  \delta_\Phi).
\ee
This property of the supercharge of our choice is important difference from
the choice in \cite{Kapustin:2009kz}, where the supercharge which is used for the localization
satisfies $Q^2=0$.
The cohomology in a usual sense is built from a nilpotent operator
itself, namely $Q^2=0$, but if we restrict observables of the forms,
on which $Q$ is acting, to ones invariant under the Lie derivative along $V$
(the $U(1)$ isometry) and the gauge transformations,
the BRST transformation (\ref{equivariant cohomology}) becomes
nilpotent on this restricted space.
In other words, if we consider only
the isometry and gauge invariant observables $\Op$ which satisfy
\be
\cL_V\Op = \delta_\Phi \Op =0,
\ee
the BRST transformations are nilpotent $Q^2\Op=0$
and $Q$ makes a cohomology on $\Op$.
Similar to the usual cohomology, one should identify an element
of the equivariant cohomology up to a $Q$-exact form
\be
\Op \sim \Op + Q \Op',
\ee
where $\Op$ satisfy $Q\Op=0$ but not $Q$-exact.
This kind of the cohomology is called equivariant cohomology.
The equivariant cohomology plays an essential role
in the proof of the localization theorem.

From the equivariant cohomology point of view, we can understand
true role of the supersymmetric CS action.
First of all, we start from 
the bosonic (non-supersymmetric) CS action, which is
\be
S_\text{CS}[A] = \frac{k}{4\pi} \int_M 
\Tr\left[
A \wedge dA -\frac{2i}{3}A \wedge A \wedge A
\right].
\label{bosonic CS}
\ee
Notice that this action is purely topological (independent of the metric on $M$)
and invariant under the gauge transformations.

Now let us consider a shift of $A$ by
\be
A \to A + i\sigma \kappa.
\ee
Recalling the definition of $\Phi$ in (\ref{definition of Phi}),
the shift of the gauge field is decomposed into
\be
A+i\sigma \kappa = A_\Sigma +  A_\kappa \kappa + i\sigma \kappa
= A_\Sigma +  \Phi \kappa,
\ee
where $A_\Sigma=A_zdz+A_\zb d\zb$
is a two-dimensional (horizontal) gauge field on $\Sigma$.
Substituting this shifted and decomposed gauge fields into
the bosonic CS action (\ref{bosonic CS}),
we find
\be
S_\text{CS}[A+i\sigma \kappa] =
\frac{k}{4\pi} \int_M
\Tr\left[
A_\Sigma \wedge \kappa \wedge \cL_V A_\Sigma
+2\Phi \kappa\wedge  F_\Sigma
+\kappa \wedge d\kappa \, \Phi^2
\right],
\label{decomposed bosonic CS}
\ee
where $F_\Sigma=dA_\Sigma -i A_\Sigma \wedge A_\Sigma$
is a two-dimensional field strength on $\Sigma$
and we have used
$A_\Sigma \wedge dA_\Sigma = A_\Sigma \wedge \kappa \wedge \cL_V A_\Sigma$.
The original bosonic CS action also has the same form
as the above (\ref{decomposed bosonic CS}) by replacing $\Phi$ by $A_\kappa$ ($\sigma=0$).
However the inclusion of the additional scalar field $\sigma$, which
is contained in the vector superfield, complexifies $A_\kappa$ to the
complex adjoint scalar field $\Phi$.
Recalling the adjoint scalar $\sigma$ originally comes from the dimensional reduction
of one component of the four-dimensional gauge field,
this complexification of $A_\kappa$ means that the circle ($S^1$) fiber bundler over $\Sigma$,
which we are considering, should be extend $M$ with (dual of) $\sigma$
to the holomorphic line bundle with the degree $p$ over $\Sigma$
\be
\begin{CD}
L @>p>> M' \\
@. @VV{\pi}V\\
 @. \Sigma
\end{CD}
\qquad ,
\ee
where $M'$ is now the four-dimensional manifold.

Secondary, we add a quadratic term of the fermions to
the shifted bosonic CS action (\ref{decomposed bosonic CS})
\be
S_\text{coh}[A,\sigma,\lambda] = S_\text{CS}[A+i\sigma \kappa] 
- \frac{k}{4\pi} \int_M 
\Tr\left[
\kappa \wedge \lambda \wedge \lambda
\right].
\ee
Then one can easily check
\be
Q S_\text{coh}[A,\sigma,\lambda] = 0,
\ee
that is, the cohomological CS action $S_\text{coh}$
is an element of the equivariant cohomology.
So $S_\text{coh}$ is a good physical observable in three-dimensional
supersymmetric gauge theory on $M$ in this sense.

To see the relation between
the cohomological CS action and
 the supersymmetric CS action,
let us rewriting the cohomological CS action to
 separate the original bosonic CS action and the residual part
including $\sigma$ and $\lambda$
\be
S_\text{coh}[A,\sigma,\lambda] = S_\text{CS}[A]
 + \frac{k}{4\pi} \int_M
\Tr\left[
2\sigma\left(
i\frac{\kappa\wedge F}{\kappa\wedge d\kappa} -\frac{1}{2}\sigma
\right)\kappa\wedge d\kappa
-\kappa \wedge \lambda \wedge \lambda
\right],
\label{cohomological action}
\ee
where we have used the fact that
\bea
\kappa \wedge F &=& \kappa \wedge \left\{dA_\Sigma + (dA_\kappa) \kappa
+A_\kappa d\kappa
-i (A_\Sigma + A_\kappa \kappa)\wedge (A_\Sigma + A_\kappa \kappa)\right\}\nn\\
&=& \kappa \wedge F_\Sigma + A_\kappa \kappa \wedge d\kappa.
\eea
Then we now add a BRST exact term 
into $S_\text{coh}$.
We see it becomes the supersymmetric CS action (\ref{SCS action})
\bea
S_\text{SCS}[A,\sigma,\lambda;Y_r,\chi_r]
&=&S_\text{coh}[A,\sigma,\lambda]+Q\frac{k}{2\pi}\int_M d^3 x\sqrt{g} \Tr[\chi_r \sigma]\nn\\
&=& S_\text{CS}[A]
 + \frac{k}{4\pi} \int_M
\Tr\Bigg[
2\sigma\left(
i\frac{\kappa\wedge F}{\kappa\wedge d\kappa}
-\frac{1}{2}\sigma+\frac{\ell}{2}Y_r
\right)\kappa\wedge d\kappa\nn\\
&&\qquad\qquad\qquad\qquad\qquad\qquad
-\kappa \wedge \lambda \wedge \lambda-i\ell\chi_r \eta \, \kappa\wedge d\kappa
\Bigg]\nn\\
&=& S_\text{CS}[A]
 + \frac{k}{4\pi} \int_M d^3 x \sqrt{g}
\Tr\left[
2 D \sigma-\tilde{\lambda}\lambda
\right],
\eea
where we have used the definition of $Y_r$. 
In summary, the difference between the cohomological CS action
and the supersymmetric action is the $Q$-exact term,
namely
$S_\text{coh}$ and $S_\text{SCS}$ belong to the same equivariant cohomology
class.
Thus
$S_\text{SCS}$ is supersymmetric ($Q$-closed) under the supercharge $Q$
\be
Q S_\text{SCS}[A,\sigma,\lambda;Y_r,\chi_r] =0.
\ee
This fact says that the additional $Q$-exact term is irrelevant when
we are considering a partition function with respect to $S_\text{SCS}$.
We can replace $S_\text{SCS}$ with $S_\text{coh}$ in the path integral
without changing the value of the partition function or vev
of the cohomological observable.


In a usual cohomological field theory,
we can use any function of $\Phi$
\be
\Op_0  =\Tr f(\Phi),
\label{trace of Phi}
\ee
as a good 0-form cohomological observable, since it
is $Q$-closed but not $Q$-exact.
However, in our formulation, the above function (\ref{trace of Phi}) itself
is not gauge invariant since $\Phi$ includes a bare gauge field $A_\kappa$.
So (\ref{trace of Phi}) is not $Q$-cohomological observable in the
supersymmetric CS theory.
A possible $Q$-closed gauge invariant observable constructed from $\Phi$ is
the Wilson loop along the $S^1$ fiber
\be
W (C) \equiv  \Tr_R \calP \exp i\oint_C  \Phi\kappa
=  \Tr_R \calP \exp i\oint_C  (A_\kappa +i \sigma ) \kappa,
\ee
where $\Tr_R$ is a trace over the representation $R$
and $\calP$ represents a path ordered product along a loop $C$,
which is oriented to the $U(1)$-fiber direction.
This Wilson loop observable is gauge invariant and does not violate the isometry
along the $U(1)$ fiber. So the BRST closed Wilson loop $W(C)$
is a good physical observable in three-dimensional cohomological
field theory on $M$.
Since $W(C)$ is nothing but the supersymmetric
Wilson loop,
the $Q$-cohomological property of $W(C)$
is a true reason why one can evaluate exactly the vev of $W(C)$
in supersymmetric gauge theory on $M$.

\section{Exact Partition Function}

\subsection{Coupling independence}

In this section, we derive the localization formula for the supersymmetric CS theory.
A similar derivation is also discussed in \cite{Bruzzo:2002xf} for the reduced
supersymmetric matrix model. We borrow some useful notations from it
in the following explanations.

So far, we have considered the supersymmetric CS action only.
If we would like to treat the supersymmetric YMCS theory,
we need to consider the supersymmetric YM action.
The supersymmetric YM action can be written as
an $Q$-exact form in general
\be
S_\text{SYM} =\frac{1}{g^2} Q\Xi,
\ee
where $\Xi$ is a functional of fields, which is invariant under the gauge symmetry
and the $U(1)$ isometry.

This $Q$-exactness of the supersymmetric YM action says that
the partition function of the supersymmetric YM theory
\be
\calZ = \int \D \Psi \, e^{-S_\text{SYM}[\Psi]},
\label{partition function}
\ee
is independent of the gauge coupling $g$,
where $\D\Psi$ is a path integral measure overall fields.
In fact, if we once differentiate the partition function with respect to the
gauge coupling, one can see
\be
\frac{\del \calZ}{\del g}
\propto
\int \D \Psi \, Q(\Xi e^{-S_\text{SYM}})=0,
\ee
since the path integral measure is made to be invariant under the
$Q$-transformation.
So we can evaluate the partition function (\ref{partition function})
in any coupling region of $g$. In particular, we can exactly
evaluate the partition function in the weak coupling region $g\to 0$.
This means that a WKB (1-loop) approximation is exact
in the evaluation of the partition function.
This 1-loop exactness of the supersymmetric gauge theory is known also as
the superrenormalizability in the perturbative expansion.

Similarly, we can show that the vev of the cohomological observable,
which satisfies $Q\Op=0$ but not $Q$-exact,
\be
\left\langle \Op \right\rangle = \calZ^{-1}
\int \D \Psi \, \Op \, e^{-S_\text{SYM}},
\ee
is also independent of the gauge coupling $g$.
So one can evaluate the vev of $\Op$ exactly in the weak coupling limit.
For example, the vev of the supersymmetric Wilson loop $W(C)$
can be evaluated exactly in the limit of $g\to0$.

As we have seen in the previous section, the supersymmetric CS action itself
is $Q$-cohomological observable. Then the supersymmetric YMCS theory
can be regarded as an evaluation of the vev of the $Q$-cohomological CS action
$\Op_\text{CS}=e^{iS_\text{SCS}}$ 
in the supersymmetric YM theory.
On the other hand, the terms containing the fermions in
the cohomological CS action is quadratic. So we can integrate out
the fermionic fields and obtain the purely bosonic (non-supersymmetric)
CS theory (plus the adjoint scalar field).
We can write these relations schematically as follows
\bea
\calZ_{\text{CS}+\sigma}&=&\int \D A \D \sigma \, e^{iS_\text{CS}[A+i\sigma \kappa]}\nn\\
&=&\int \D A \D \sigma \D\lambda \, e^{iS_\text{coh}[A,\sigma,\lambda]}
=\calZ_\text{SCS}\nn\\
&=&\left\langle
e^{iS_\text{SCS}}
\right\rangle_\text{SYM}
=
\calZ_\text{SYMCS},
\eea
by using the gauge coupling independence, 
where $\langle \cdots \rangle_\text{SYM}$
stands for the vev in the supersymmetric 
YM theory.
This is the reason why the partition functions of the supersymmetric YMCS theories
are the same as those of the bosonic (non-supersymmetric) CS theories.\footnote{
As we will see, there is a discrepancy in the CS level between the bosonic
and supersymmetric theories. We have here ignored the quantum corrections
(anomalies) from the integration of the chiral fermions.}


\subsection{The localization}

To proceed a proof of the localization theorem in the supersymmetric YM theory,
we explicitly give the $Q$-exact action by
\be
S_\text{SYM} = \frac{1}{2g^2}Q \int_M
\Tr\left[
\vec{\F}_v \wedge \star \overline{Q\vec{\F}_v}- i\ell\chi_r \mu_r \, \kappa\wedge d\kappa
\right],
\label{super YM}
\ee
where $\wedge \star$ represents suitable norms
among different degree of the field forms,
including the inner product of the vector. 
We here combine the bosonic and fermionic fields into vectors
$\vec{\B}_v \equiv (A_\Sigma,\Phib,Y_r)$ and
$\vec{\F}_v \equiv (\lambda_\Sigma,\eta,\chi_r)$,
where $\Phib \equiv A_\kappa -i\sigma$.
We have decomposed the form of fields into the components on the 
horizontal Riemann surface $\Sigma$ and ones along the circle fiber.
Since the adjoint scalar $\Phi$ is $Q$-closed itself and does not have
fermionic partner, we should treat this combination of fields rather special.
$\Phi$ will play an important role in the localization of the path integral.
$\mu_r$ in (\ref{super YM}) defined by
\be
\mu_r \equiv  \frac{\kappa\wedge F}{\kappa\wedge d\kappa}
\ee
corresponds to the D-term constraint in
the supersymmetric YM theory. It is called 
the moment map (of the \Kahler quotient space).
Roughly speaking, the partition function of the YM theory on $M$ measures the volume of the
flat connection moduli space on $M$ $\Vol({\cal M}_{F=0})$, where
${\cal M}_{F=0}=\mu_r^{-1}(0)/G$ is the \Kahler quotient space.

In the above decomposition of the fields, the BRST transformations become
\be
\begin{array}{ll}
Q A_\Sigma = \lambda_\Sigma, & Q \lambda_\Sigma = -i(\cL_V A_\Sigma
-i[\Phi,A_\Sigma]-\pi^* d_\Sigma \Phi),\\
Q \Phi =0, &\\
Q \Phib = 2\eta, &Q\eta = -\frac{i}{2}\left(\cL_V \Phib-i[\Phi,\Phib]-\cL_V\Phi\right),\\
Q Y_r =-i(\cL_V \chi_r-i[\Phi,\chi_r]), & Q \chi_r =Y_r,
\end{array}
\ee
where $\pi^* d_\Sigma$ is the pullback of an exterior derivative restricted on $\Sigma$.
$\overline{Q\vec{\F}_v}$ is regarded as hermite conjugate of these
transformations.
Using these BRST transformations, we find that the bosonic part of
the above $Q$-exact action is 
\bea
\left. S_\text{SYM}  \right|_\text{boson}
&=&\frac{1}{2g^2} \int_M
\Tr\left[
Q\vec{\F}_v \wedge \star \overline{Q\vec{\F}_v}-i \ell Y_r \mu_r \kappa\wedge d\kappa
\right]\nn\\
&=&\frac{1}{2g^2} \int_M
\Tr\left[
d_A\sigma \wedge \star d_A\sigma +\iota_V F \wedge \star \iota_V F
+\frac{\ell}{2}Y_r(Y_r -2i \mu_r)\kappa\wedge d\kappa
\right].\nn\\
\eea
After integrating out the auxiliary field $Y_r$, the action reduces to
\bea
\left. S_\text{SYM}  \right|_\text{boson}
&=&\frac{1}{2g^2} \int_M
\Tr\left[
d_A\sigma \wedge \star d_A\sigma +\iota_V F \wedge \star \iota_V F
+\frac{\ell}{2}\mu_r^2 \, \kappa\wedge d\kappa
\right]\nn\\
&=&\frac{1}{2g^2} \int_M d^3x \sqrt{g}
\Tr\left[\frac{1}{2} F_{\mu\nu}F^{\mu\nu}
+\D_\mu \sigma \D^\mu \sigma
\right].
\eea
This is exactly the bosonic part of
the supersymmetric YM action (\ref{SYM action}) obtained by integrating out
the auxiliary D-field.
We can see the fermionic part of $S_\text{SYM}$
 also agrees with the supersymmetric YM one
 up to the field redefinitions defined in the previous section.


Let us now discuss the meanings of the supersymmetric YM action
(\ref{super YM}).
To make clearer the role of each term in (\ref{super YM}),
we separate the terms by adding the extra parameter $t_1$ and $t_2$
\be
S_\text{SYM} = \frac{1}{2g^2}Q \int_M
\Tr\left[
t_1 \left(\vec{\F}_v \wedge \star \overline{Q\vec{\F}_v}\right)
-t_2\left( 2i\chi_r \mu_r \, \kappa\wedge d\kappa\right)
\right]
\ee
Since both terms are still $Q$-exact independently, the partition function
is independent of the parameters $t_1$ and $t_2$.

If we consider a situation that $t_1\ll t_2$, the bosonic part proportional to $t_2$
gives a delta-functional constraint at $\mu_r=0$. And also, the fermionic part
will give suitable Jacobians for the constraint of $\mu_r=0$.
Thus the terms proportional to $t_2$ of the supersymmetric YM action
give the D-term constraint $\delta(\mu_r)$ with the suitable Jacobians
in the path integral of the partition function.

In the other limit of $t_1\gg t_2$, the bosonic part proportional to $t_1$
is essentially Gaussian of the field variables $Q\vec{\F}_v$.
In the $t_1 \to \infty$ limit, the Gaussian integral approximates
to the delta-functional at $Q\vec{\F}_v=0$, namely
the path integral is localized at the fixed point set of the
BRST transformations $Q\vec{\F}_v=0$.
This is the mechanism of the localization in the supersymmetric gauge theory.
The fermionic part proportional to $t_1$ cancels out the $t_1$ dependence
of the Gaussian integral of the bosonic part
(and gives the fixed point at $\vec{\F}_v=0$) since the degree of freedom
of the bosonic and fermionic fields are the same.
Both integral over the bosonic and fermionic fields give some Jacobians
(1-loop) determinants into the path integral. We need to
evaluate these Jacobians to obtain the localization formula.

In summary, the path integral localizes at the fixed points $Q\vec{\F}_v =0$ on 
the D-term constraint $\mu_r=0$.
Precisely speaking, the existence of $\bar{\Phi}$,
which complexifies the gauge transformation,
relaxes the D-term constraint $\mu_r=0$
to admit the higher critical points $\mu_r\neq 0$ \cite{Witten:1992xu}.
It reflects the fact that
the \Kahler quotient by the gauge group $\mu_r^{-1}(0)/G$
is equivalent to a quotient by the complexified gauge group $G_\C$
without the D-term (real moment map) constraint.
The D-term constraint can not be imposed until the BF-type action
$S_\text{BF}=i\Tr \int \Phi \mu_r$ is inserted.
After integrating out $\Phi$ completely, we get the localization on the quotient space
of the flat connections ${\cal M}_{F=0}$.
However, for two dimensional YM theory or three dimensional CS theory,
there exist the contributions from the higher critical points which contain
not only $\mu_r=0$ but also $\mu_r \neq 0$ because of the quadratic potential of $\Phi$.

\subsection{1-loop determinants}

Now let us evaluate the Jacobians (1-loop determinants) precisely
in the limit of $t_1\to \infty$,
as follows.
Similar arguments are discussed in \cite{Karki:1993bw} for the classical Hamiltonian system.

First of all, we expand the fields around the fixed point set by
\be
\B^I_v = \B^I_{v,0} + \frac{1}{\sqrt{t_1}}\tilde{\B}^I_v,
\qquad
\F^I_v = 0 + \frac{1}{\sqrt{t_1}}\tilde{\F}^I_v,
\ee
where $\B^I_v$ and $\F^I_v$ are each components of
$\vec{\B}_v$ and $\vec{\F}_v$, and
$\B^I_{v,0}$ are solutions to $Q\vec{\F}_v=0$.
Substituting this expansion into the Gaussian part of the supersymmetric YM
action, it becomes
\bea
S'_\text{SYM} &=& \frac{t_1}{2g^2} \int_M
\Tr\left[
Q \F^I_v \wedge \star \overline{Q\F_{vI}}
-\F^I_v \wedge \star Q( \overline{Q\F_{vI}})
\right]\nn\\
&=&
\frac{1}{2g^2} \int_M
\Tr\left[
G_{IJ}
\tilde{\B}^I_v \wedge \star \tilde{\B}^J_v
+\frac{1}{2}\Omega_{IJ}\tilde{\F}^I_v \wedge \star \tilde{\F}^J_v
\right]+\Op(1/t_1),
\label{expansion}
\eea
where
\bea
G_{IJ} &=&\left.
\frac{\delta^2}{\delta \B_v^I \delta \B_v^J}
\left(
Q \F^K_v \wedge \star \overline{Q\F_{vK}}
\right)
\right|_{\vec{\B}_v=\vec{\B}_{v,0}},\\
\Omega_{IJ} &=&
\left.
\frac{\delta ( \overline{Q\F_{vI}} )}{\delta \B^{J}_v}
-\frac{\delta ( \overline{Q\F_{vJ}} )}{\delta \B^{I}_v}
\right|_{\vec{\B}_v=\vec{\B}_{v,0}}.
\eea
In the $t_1\to \infty$ limit, the Gaussian integrals
of both bosonic and fermionic parts go to the delta-functionals
\bea
&&\lim_{t_1\to \infty}
e^{-\frac{t_1}{2g^2}\int_M
\Tr
Q \F^I_v \wedge \star \overline{Q\F_{vI}}}
=\left(\frac{1}{\pi g^2}\right)^{-n_B/2} \frac{1}{\sqrt{\Det |G_{IJ}|}}
\delta(\vec{\tilde{\B}}_v),\\
&&\lim_{t_1\to \infty}
e^{\frac{t_1}{2g^2}\int_M
\Tr
\F^I_v \wedge \star Q( \overline{Q\F_{vI}})}
=\left(\frac{1}{g^2}\right)^{n_F/2} \sqrt{\Det |\Omega_{IJ}|}
\delta(\vec{\tilde{\F}}_v),
\eea
where the determinants are taken overall
modes and representations of the fields, and
$n_B$ and $n_F$ are the total number of the bosonic
and fermionic modes, respectively.  
Combining these terms together,
we obtain the exact result (up to irrelevant infinite constants
which can be absorbed into the path integral measure)
\be
e^{-S'_\text{SYM}}
=\sqrt{\frac{\Det |\Omega_{IJ}|}{\Det |G_{IJ}|}}
\delta(Q\vec{\F}_v)\delta(\vec{\F}_v).
\label{Jacobians}
\ee
which is independent of $t_1$ and $g$ as expected,  since $n_B=n_F$
because of the supersymmetry.
Thus the path integral of the supersymmetric gauge theory has 
supports only on the fixed point set $Q\vec{\F}_v=\vec{\F}_v=0$
with the Jacobians.

Since the supersymmetric YM action is $Q$-exact, it is obviously satisfied
that $QS'_\text{SYM}=0$. This $Q$-closedness must be satisfied 
in any order of $t_1$ in the expansion (\ref{expansion}).
In particular, from the $Q$-closedness of the leading term in (\ref{expansion}),
we find that\footnote{
We can  show the same relation from the Killing equation
for the metric $G_{IJ}$.}
\be
G_{IK}
\frac{\delta (Q\B_v^K)}{\delta \F_v^J}
=\Omega_{IK}\frac{\delta (Q\F_v^K)}{\delta \B_v^J}.
\ee
Then we get
\be
\Det |G_{IJ}|
\Det\left|\frac{\delta (Q\B_v^I)}{\delta \F_v^J}\right|
=\Det|\Omega_{IJ}|
\Det\left|\frac{\delta (Q\F_v^I)}{\delta \B_v^J}\right|,
\ee
at $Q\vec{\F}_v=\vec{\F}_v=0$.
Substituting this into (\ref{Jacobians}), we finally obtain
\be
e^{-S'_\text{SYM}}
=\sqrt{
\frac{
\Det\left|\frac{\delta (Q\B_v^I)}{\delta \F_v^J}\right|}
{\Det\left|\frac{\delta (Q\F_v^I)}{\delta \B_v^J}\right|}}
\delta(Q\vec{\F}_v)\delta(\vec{\F}_v).
\label{1-loop determinants}
\ee
Thus the Jacobians (1-loop determinants) are
represented only in terms of the field differentials
of the BRST transformations.

\subsection{Fixed points and localization formula}

We have seen that the path integral of the supersymmetric YM theory
localizes at the fixed point equation $Q\vec{\F}_v=0$ of the
BRST transformations.
In order to solve this BRST fixed point equations more explicitly,
we here decompose all fields in the adjoint representation as
follows
\be
\Psi = \sum_{a=1}^r \Psi^a H_a 
+ \sum_{\alpha>0} \Psi^\alpha E_\alpha
+ \sum_{\alpha>0} \Psi^{-\alpha} E_{-\alpha},
\ee
where $H_a$'s ($a=1,\ldots, r=\rank \mathfrak{g})$
are  generators of the Cartan subalgebra
of Lie algebra $\mathfrak{g}$
and $E_\alpha$ and $E_{-\alpha}$ are  generators
associated with the root $\alpha$.
The generators satisfy the following relations
\bea
&&[H_a,H_b]=0,\quad [H_a,E_\alpha]=\alpha_a E_\alpha,\\
&& \Tr E_\alpha E_\beta = \delta_{\alpha+\beta,0}.
\eea

Since the supersymmetric YM theory has the non-Abelian
gauge symmetry $G$, we need to fix the gauge.
Firstly, we choose a gauge which ``diagonalizes'' $\Phi$
by setting $\Phi^\alpha=\Phi^{-\alpha}=0$, namely
\be
\Phi = \sum_{a=1}^r \Phi^a H_a,
\ee
in this gauge (but $\Phib$ is not diagonalized). 
Under this choice of the gauge, there still remains Abelian gauge
groups associated with the Cartan subalgebra.
So we secondly require a gauge fix condition $\cL_V A^a=0$
for the Abelian part of the gauge fields.
These gauge conditions fix all of the gauge symmetry of $G$
and induce additional functional determinants in the path integral
from an action for ghosts $c$ and $\bar{c}$
\be
S_\text{ghost}[c,\bar{c}] = \frac{1}{2g^2} \int_M
d^3x \sqrt{g}\,
\Tr\left[
c(\cL_V \bar{c} -i[\Phi,\bar{c}])
\right]
.
\ee

Using the above gauge fixing condition, the BRST fixed point equation
$Q\vec{\F}_v=0$ says
\bea
&&\left(\cL_V -i\alpha(\Phi)\right) A^\alpha_\Sigma
=\left(\cL_V + i\alpha(\Phi)\right) A^{-\alpha}_\Sigma
=0,\\
&&\left(\cL_V -i\alpha(\Phi)\right) \Phib^\alpha_\Sigma
=\left(\cL_V + i\alpha(\Phi)\right) \Phib^{-\alpha}_\Sigma
=0,\\
&& d\Phi =0,
\label{dPhi=0}\\
&&Y_r=0,
\eea
where $\alpha(\Phi)\equiv \sum_{a=1}^r \alpha_a \Phi^a$.
The third line of the fixed point equations (\ref{dPhi=0}) means
that $\Phi^a(x)$ is constant everywhere on $M$. 
We denote these constant zero modes by $\phi_a$ in the following.

After integrating out all off-diagonal components of the adjoint fields,
including the ghosts,
with taking care on the 1-loop determinants (\ref{1-loop determinants}),
we obtain the partition function of
the Abelian gauge theory as a result of the localization
\be
\calZ_\text{SYM} = 
\frac{1}{|W|} \int \prod_{a=1}^r
\left\{
d \phi_a  \D \A^a_{\Sigma}\D \lambda^a
\right\}
\prod_{\alpha\neq 0}
\frac{\Det_{c,\bar{c}}|\cL_V -i\alpha(\phi)|}
{\Det_{A_\Sigma}|\cL_V -i\alpha(\phi)|}
\sqrt{
\frac{\Det_{\chi_r}|\cL_V -i\alpha(\phi)|}
{\Det_{\Phib}|\cL_V -i\alpha(\phi)|}
},
\ee
where $|W|$ is the order of the Weyl group of $G$,
and we have assigned the field subscripts of the determinant
in order to show which non-zero modes of the fields
 the determinant is taken over.

Since the number of modes of $\Phib$ and $\chi_r$ (0-forms) are the same,
the determinants of $\Phib$ and $\chi_r$ are cancelled with each other.
This reflects the fact that the D-term constraint is absorbed into
the complexification of the gauge group by $\Phib$.
The partition function becomes simply
\be
\calZ_\text{SYM} = 
\frac{1}{|W|} \int
\prod_{a=1}^r
\left\{
d \phi_a  \D \A^a_{\Sigma}\D \lambda^a
\right\}
\prod_{\alpha\neq 0}
\frac{\Det_{c,\bar{c}}|\cL_V -i\alpha(\phi)|}
{\Det_{A_\Sigma}|\cL_V -i\alpha(\phi)|}.
\ee
Note here that the determinants are definitely the absolute value,
so there is no phases from the determinants,
because of the hermiticity (no oscillatory nature)
and the invariance under the parity symmetry, $\Phi\to -\Phi$ and $\Phib\to -\Phib$
of the supersymmetric YM action. 

We now expand the fields on $M$ by
\be
\Psi(z,\zb,\theta) = \sum_{n\in\Z} \Psi_n(z,\zb)e^{-i n \theta/\ell},
\ee
where $\Psi_n(z,\zb)$, which satisfy $\cL_V\Psi_n=0$, are functions on $\Sigma$ and
we denote $\theta$ as the coordinate along the circle fiber direction.
Then, for each $n$, the eigenvalue of $\cL_V$ on $\Psi_n$ is given by $-i\frac{n}{\ell}$,
which means that $\Psi_n$ are sections of line bundles $\Op(-pn )$ over $\Sigma$.
The difference of the number of those modes between
0-forms ($c,\bar{c}$) and 1-form $A_\Sigma$ on
$\Sigma$ is given by the Hirzebruch-Riemann-Roch theorem
\be
\dim \Omega^0(\Sigma,\Op(-pn)\otimes V_\alpha)
-\dim \Omega^1(\Sigma,\Op(-pn) \otimes V_\alpha)
=\frac{1}{2}\chi(\Sigma)-pn+\alpha(m),
\ee
where
$\chi(\Sigma)=2-2h$ is the Euler number of the Riemann surface.
The second term comes from the first Chern class $c_1(\Op(-pn))$.
The third term
$\alpha(m)=\sum_{a=1}^r \alpha_a m_a$ is made from
the first Chern class (magnetic flux) of the
$a$-th background $U(1)$ gauge field on $\Sigma$, namely
$m_a=\frac{1}{2\pi}\int_\Sigma F_a$.

Using the above observations,
the localization formula for the partition function reduces further to
\be
\calZ_\text{SYM} = 
\frac{1}{|W|} \int \prod_{a=1}^r
\left\{
d \phi_a \D \A^a_{\Sigma}\D \lambda^a
\right\}
\prod_{n\in \Z}
\prod_{\alpha \neq 0}
\left|
\frac{n}{\ell} + \alpha(\phi)
\right|^{\frac{1}{2}\chi(\Sigma)-pn+\alpha(m)}.
\ee
We now evaluate the infinite product in the above integrand.
We first see
\bea
\prod_{n\in \Z}
\prod_{\alpha \neq 0}
\left|
\frac{n}{\ell} + \alpha(\phi)
\right|^{\frac{1}{2}\chi(\Sigma)-pn+\alpha(m)}
&=&
\prod_{n\in \Z}
\prod_{\alpha>0}
\left(
\frac{n}{\ell} + \alpha(\phi)
\right)^{\chi(\Sigma)}\nn\\
&=&
\prod_{\alpha>0}
\alpha(\phi)^{\chi(\Sigma)}
\prod_{n=1}^\infty
\left\{
\frac{n^2}{\ell^2}
\left(
1-\frac{\ell^2\alpha(\phi)^2}{n^2}
\right)
\right\}^{\chi(\Sigma)}\nn\\
&=&
\prod_{\alpha>0}
\left(
2\sin \pi \ell \alpha(\phi)
\right)^{\chi(\Sigma)},
\label{1-loop determinant}
\eea
where we have used the infinite product expansion of the sin-function
\be
\sin(\pi z) = {\pi z}\prod_{n=1}^\infty\left(
1-\frac{z^2}{n^2}
\right),
\ee
and the zeta-function regularization of the infinite product
\be
\prod_{n=1}^\infty
\frac{n^2}{\ell^2}
=e^{
-2\zeta'(0)-2 \zeta(0) \log \ell
}
=2\pi \ell.
\ee
Note that the dependence on the $U(1)$ gauge fields is disappeared from the determinant here.

We also would like to emphasis here that there is no phase coming from the infinite product of signs in particular.  
In the bosonic CS case, where the determinants are not the absolute value,
the phase from the determinants is evaluated explicitly and leads to the famous quantum shift in the CS level
$k \to k+ \check{c}_\mathfrak{g}$, where $\check{c}_\mathfrak{g}$ is the dual Coxeter number of $\mathfrak{g}$.
Our claim is that there is no level shift in the  $\N=2$ {\it supersymmetric} YMCS theory.
This is consistent with the following physical consideration \cite{Kao:1995gf}:
In the $\N=2$ supersymmetric gauge theory, there are two chiral fermions $\lambda,\tilde{\lambda}$
with the same chirality
in the adjoint representation.
These chiral fermions reproduce the CS action due to the parity anomaly
in three-dimensions. Each chiral fermion contribute to the CS level by $\pm\frac{1}{2}\check{c}_\mathfrak{g}$,
where the sign depends on the chirality of the fermions relative to the bare CS level.
Then the total level shift in the ${\cal N}=2$ supersymmetric CS theory is given by
$k\to k +  \check{c}_\mathfrak{g} -\frac{1}{2} \check{c}_\mathfrak{g}-\frac{1}{2} \check{c}_\mathfrak{g} = k$,
namely there is no quantum level shift.
Incidentally, the quantum level shift in the ${\cal N}=1$ and ${\cal N}=3$ supersymmetric CS theory
is $k\to k+\frac{1}{2} \check{c}_\mathfrak{g}$ and $k\to k$, respectively.

Since YM part is $Q$-exact and coupling independent,
 it seems that the limit $g^2 \to \infty$ does not  change the results and 
the YMCS theory is equivalent to  pure CS theory with free auxiliary gaugino in which
the quantum level shift actually occurs \cite{Thompson:2010}. 
One might think this conflicts with the above result.
But as discussed in \cite{Tanaka:2012nr}, since
the limit of $g^2 \to \infty $ and the path integral for gaugino do not commute,   
the supersymmetric YMCS theory on the Seifert manifold does not receive the 
level shift in the contrast to pure CS theory. 
We will later calculate the supersymmetric indexes by the localization and
see agreement with the results predicted by brane construction for 
supersymmetric YMCS theories. This confirm that no level shift occurs in our formulation.

Finally we obtain the localization formula for the supersymmetric YM theory on
$M$
\be
\calZ_\text{SYM} = 
\frac{1}{\ell^r|W|} \int_{-\infty}^\infty \prod_{a=1}^r
d \phi_a
\prod_{\alpha>0}
\left(
2\sin \frac{\alpha(\phi)}{2}
\right)^{\chi(\Sigma)},
\ee
where we have used the normalization
of the volume of the Abelian
gauge group $H=U(1)^r$ associated with the Cartan subalgebra
\be
\frac{1}{(2\pi)^r}\int \prod_{a=1}^r \D A^a_{\Sigma}\D \lambda^a=\frac{\Vol(H)}{(2\pi)^r}=1,
\ee
and rescaled $2\pi\ell \phi_a \to \phi_a$ to be the integral over the dimensionless variables.
The $U(1)$ integrals can be factored out since the integrand is independent of
the background $U(1)$ fields.

\subsection{The supersymmetric YMCS partition function}

We arrived at the evaluation of the partition function of the supersymmetric YMCS theory at last.
As we have explained above, the inclusion of the CS action does not change the localization fixed points
and the derived 1-loop determinants since $e^{iS_\text{CS}}$ is the $Q$-closed cohomological observable.
We should just evaluate $e^{iS_\text{CS}}$, or equivalently $e^{iS_\text{coh}}$ at the fixed points.

At the localization locus, the cohomological CS action reduces to
\be
\left.S_\text{coh}\right|_\text{fixed points}
=k\ell\sum_{a=1}^r \int_\Sigma \Tr\left[
\phi_a  F_{\Sigma}^a
+\frac{p}{4\ell}\phi_a^2 \, \omega
- \frac{1}{2} \lambda^a \wedge \lambda^a
\right],
\label{CS at fixed point}
\ee
where we have also used the gauge fixing condition $\cL_V A^a=0$.
This action is exactly the same as the two-dimensional YM theory one (two-dimensional cohomologial
BF theory plus the mass term)!
Thus we see that the supersymmetric YMCS theory on the Seifert manifold $M$
is almost the same as the supersymmetric (cohomological) YM theory on $\Sigma$
except for the 1-loop determinant (\ref{1-loop determinant}).
Using $\phi_a$ is constant on $\Sigma$ and $\lambda^a=0$ at the fixed points, 
(\ref{CS at fixed point}) reduces further to
\be
\left.S_\text{coh}\right|_\text{fixed points}
= k\sum_{a=1}^r \left[
2\pi \ell \phi_a m_a+\frac{p\A}{4}\phi_a^2
\right],
\ee
with the $U(1)$-fluxes $m_a =\frac{1}{2\pi}\int_\Sigma F^a_\Sigma$.

Combining them together, we finally obtain the 
matrix model like integral formula for the partition function of
the supersymmetric YMCS theory on $M$
\be
\calZ_\text{SYMCS}
 = \frac{1}{\ell^r|W|}\sum_{\vec{m}\in (\Z_p)^r}
\int_{-\infty}^\infty \prod_{a=1}^r \frac{d\phi_a}{2\pi}
\prod_{\alpha>0}
\left(
2\sin 
 \frac{\alpha(\phi)}{2}
\right)^{\chi(\Sigma)}
e^{i k \sum_{a=1}^r \left[\phi_a m_a+\frac{p\mu}{4\pi} \phi_a^2\right]},
\label{phi integral of SYMCS}
\ee
where we have rescaled $2\pi \ell \phi_a \to \phi_a$
and $\mu\equiv \frac{\pi \A}{(2\pi\ell)^2}$
is a dimensionless parameter, which measures the ratio of the size of the base $\Sigma$ to the $S^1$ fiber.
The summation of $\vec{m}=(m_1,m_2,\ldots,m_r)$ is taken over
the topological sectors of the $U(1)$-fluxes admitted in this theory.
On the Seifert manifold $M$, the first Chern class on $\Sigma$ is
restricted in $\Z_p$, namely $m_a = 0,1,\ldots,p-1$, since
$H^2(M,\Z)=H^1(\Sigma,\Z)\oplus \Z_p$ \cite{Blau:2006gh}.
This integral expression agrees with the CS matrix integral on $S^3$ \cite{Marino:2002fk,Aganagic:2002wv}
by changing the integral contour.

Let us investigate the matrix integral (\ref{phi integral of SYMCS}) further.
If we consider the limit of $p\to \infty$ with $p\mu$ fixed, 
the summation over $\vec{m}$ is unrestricted.
Using the Poisson resummation of the periodic delta function
\be
\frac{1}{(2\pi)^r}\sum_{\vec{m}\in \Z^r} \prod_{a=1}^r e^{i k  \phi_a m_a}
=\frac{1}{k^r} \sum_{\vec{n}\in \Z^r} \prod_{a=1}^r \delta\left(\phi_a -\frac{2\pi n_a}{k}\right),
\label{periodic delta function}
\ee
we can integrate out $\phi_a$'s in (\ref{phi integral of SYMCS}) explicitly, then the partition function
of the supersymmetric YMCS is expressed in terms of the summation only over the discrete integer set
\be
\calZ_\text{SYMCS}
 = \frac{1}{(\ell k)^r|W|}\sum_{\vec{n}\in \Z^r}
\prod_{\alpha>0}
\left(
2\sin 
\frac{\pi}{k} \alpha(n)
\right)^{\chi(\Sigma)}
e^{i \frac{\pi}{k}p\mu \sum_{a=1}^r n_a^2 },
\ee
where $\alpha(n)=\sum_{a=1}^r \alpha_a n_a$.

This partition function is exactly the same as the $q$-deformed two-dimensional YM theory one
\cite{Aganagic:2004js}
by the identification
\be
g_s \equiv -\frac{2\pi i}{k}.
\ee
To see this more explicitly, we consider the case of $G=U(N)$ in the following.
The partition function becomes
\be
\calZ_\text{SYMCS}
 = C_N
\sum_{n_1>\cdots>n_N}
\prod_{1\leq a < b \leq N}
\left[
n_a-n_b
\right]_q^{\chi(\Sigma)}
q^{\frac{1}{2}p\mu \sum_{a=1}^N n_a^2 },
\ee
where $C_N \equiv e^{i\frac{3}{4}\pi \chi(\Sigma) N(N-1)}
 \left(\frac{ig_s}{\pi\ell}\right)^N$
 and $[x]_q$ is the $q$-number
\be
[x]_q \equiv q^{x/2} - q^{-x/2},
\ee
with 
$q \equiv e^{-g_s} = e^{\frac{2\pi i}{k}}$.
Here we have replaced the sum over the integer set $n_a\in \Z^N$
with an non-colliding integer set $n_1>n_2>\cdots>n_N$ with a fixed order using the Weyl group,
since $[n_a-n_b]_q$ vanishes and drops from the sum of the partition function if $n_a=n_b$.

These non-colliding integer set can be also expressed by the ordered colliding integer set
$\nu_1 \geq \nu_2 \geq \cdots \geq \nu_N$ through
\be
n_a \equiv \nu_a -a +\frac{N+1}{2},
\ee
that is, $\nu_a$ can be identified with the number of $a$-th row of the Young diagram.
Then the summation over $n_a$ is equivalent to the summation over the representation $R$
associated with the Young diagram $Y=(\nu_1,\nu_2,\ldots,\nu_N)$.
Thus we obtain
\be
\frac{\calZ_\text{SYMCS}}{\calZ_0}
 =
\sum_{R}
(\dim_q R)^{\chi(\Sigma)}
q^{\frac{1}{2}p\mu C_2(R) },
\ee
where
\be
\dim_q R = \prod_{1\leq a < b \leq N} \frac{[\nu_a-\nu_b+b-a]_q}{[b-a]_q},
\ee
is the quantum dimension of the representation $R$ and
\be
C_2(R) = \kappa_R + N|R|,
\ee
is the quadratic Casimir in the representation $R$
with
$\kappa_R \equiv \sum_{a=1}^N \nu_a(\nu_a-2a+1)$ and
$|R| \equiv \sum_{a=1}^N \nu_a$.

Here $\calZ_0$ is the contribution to the partition function from the lowest critical point (``ground state'')
\be
\calZ_0 =  C_N \prod_{1\leq a < b \leq N} [b-a]_q^{\chi(\Sigma)} q^{p\mu\frac{N(N^2-1)}{24}},
\ee
where  the Young diagram is empty 
or $n_a$ belongs to the Weyl vector of $SU(N)$,
namely $n_a=\rho_a\equiv \frac{N+1}{2} -a$.
$\calZ_0$ is essentially the partition function of the purely bosonic CS theory on $M$ \cite{Witten:1988hf}
without the quantum level shift up to some overall constants.
We discuss a meaning of the choice of various partitions next.

\subsection{The supersymmetric index}

The other case where we can use the resummation formula is the $p=0$ case,
namely $M$ is the direct product $M=S^1\times \Sigma$.
In this case, there is no torsion $M$ and
the second cohomology on $\Sigma$ is not restricted, then
$H^2(\Sigma,\Z)=\Z$.
Using the periodic delta function (\ref{periodic delta function}),
the partition function of the $U(N)$ gauge theory can be evaluated by
\be
\calZ_\text{SYMCS}(S^1\times \Sigma)
 = C_N
\sum_{n_1>\cdots>n_N}
\prod_{1\leq a < b \leq N}
\left[
n_a-n_b
\right]_q^{\chi(\Sigma)},
\ee
which could be called the partition function of 
 the $q$-deformed two-dimensional BF theory on $\Sigma$.

The summation here is taken over the ordered non-colliding integer set, but 
we must be careful $q$ is at the root of unity $e^{\frac{2\pi i}{k}}$ in the CS theory.
Then the $q$-number $([x]_q)^{\chi(\Sigma)}$ for even $\chi(\Sigma)$ becomes periodic (sine-function) and invariant under the shift
$x\to x+ k s$ ($s\in \Z$).
So if we define
\be
n_a = k s_a + \tilde{n}_a,
\ee
where $s_a\in \Z$ and $\tilde{n}_a = 0,1,\ldots,k-1$, which are chosen to
satisfy $n_1>n_2>\cdots>n_N$,
then the partition function is written in terms of
 the summation over the fundamental domain of $\tilde{n}_a$
\be
\calZ_\text{SYMCS}(S^1\times \Sigma)
 = \tilde{C}_N
\sum_{\tilde{n}_1>\cdots>\tilde{n}_N}
\prod_{1\leq a < b \leq N}
\left[
\tilde{n}_a-\tilde{n}_b
\right]_q^{\chi(\Sigma)},
\ee
where the constant $\tilde{C}_N \equiv C_N  \left(\frac{1}{N!}
\sum_{\vec{s} \in \Z^N}1
 \right)$ is regularized by the zeta function for example.

Note that 
there is no choice of the summation and the partition function
vanishes if $N>k$,
since the summation is taken over
$N$ integers $\tilde{n}_a$, which are chosen within the integers from $0$ to $k-1$ 
with the non-colliding order $\tilde{n}_1>\cdots>\tilde{n}_N$.
This fact strongly suggests that the supersymmetric index (Witten index) 
\be
{\cal I} = \Tr (-1)^F
\ee
also
vanishes and the supersymmetry may be dynamically broken if $N>k$.
It agrees with the discussion in \cite{Witten:1999ds}
and the string theoretical explanation in \cite{Bergman:1999na,Ohta:1999iv}.

\begin{figure}[t]
\begin{center}
\begin{tabular}{ccc}
\includegraphics[scale=0.35]{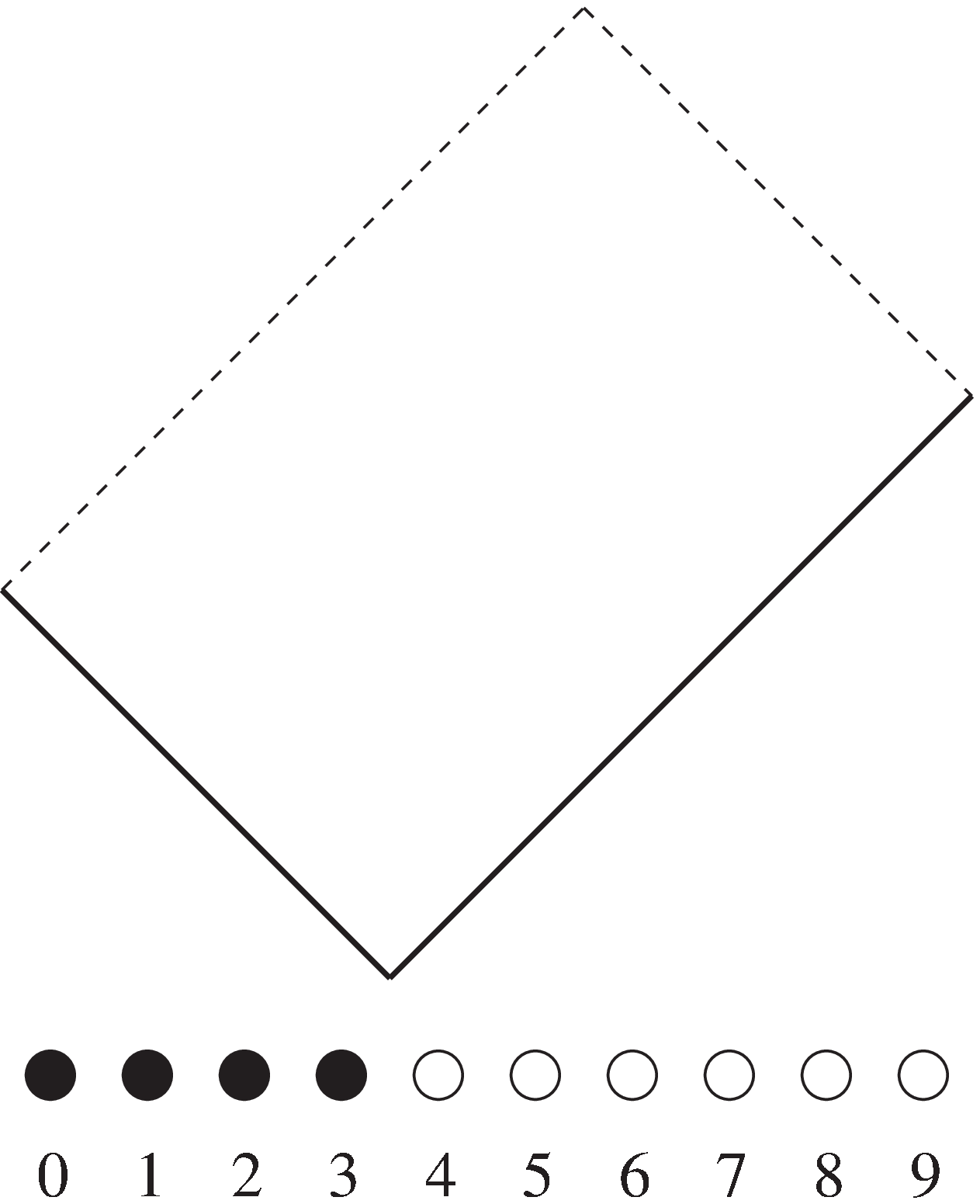}&
\includegraphics[scale=0.35]{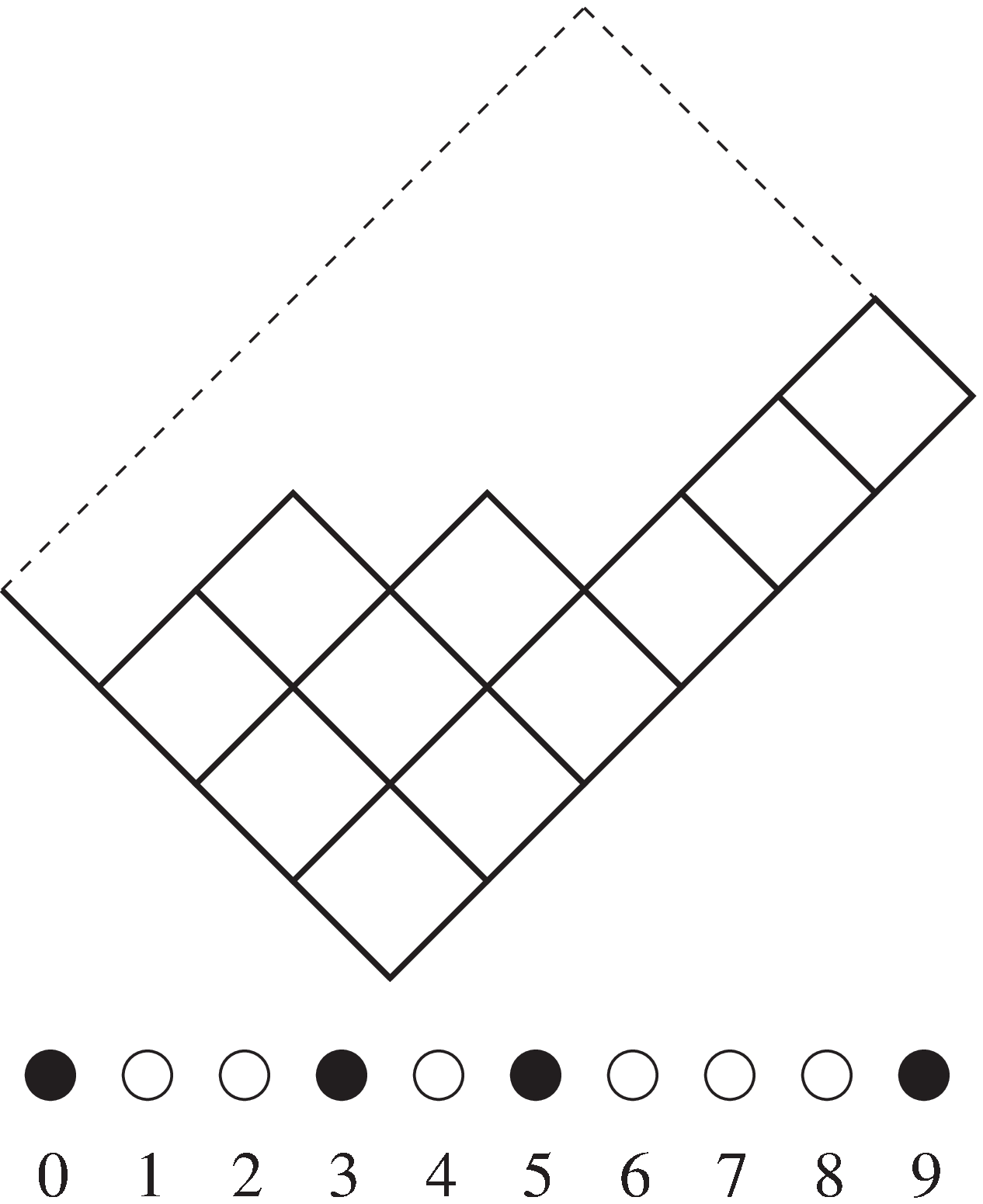}&
\includegraphics[scale=0.35]{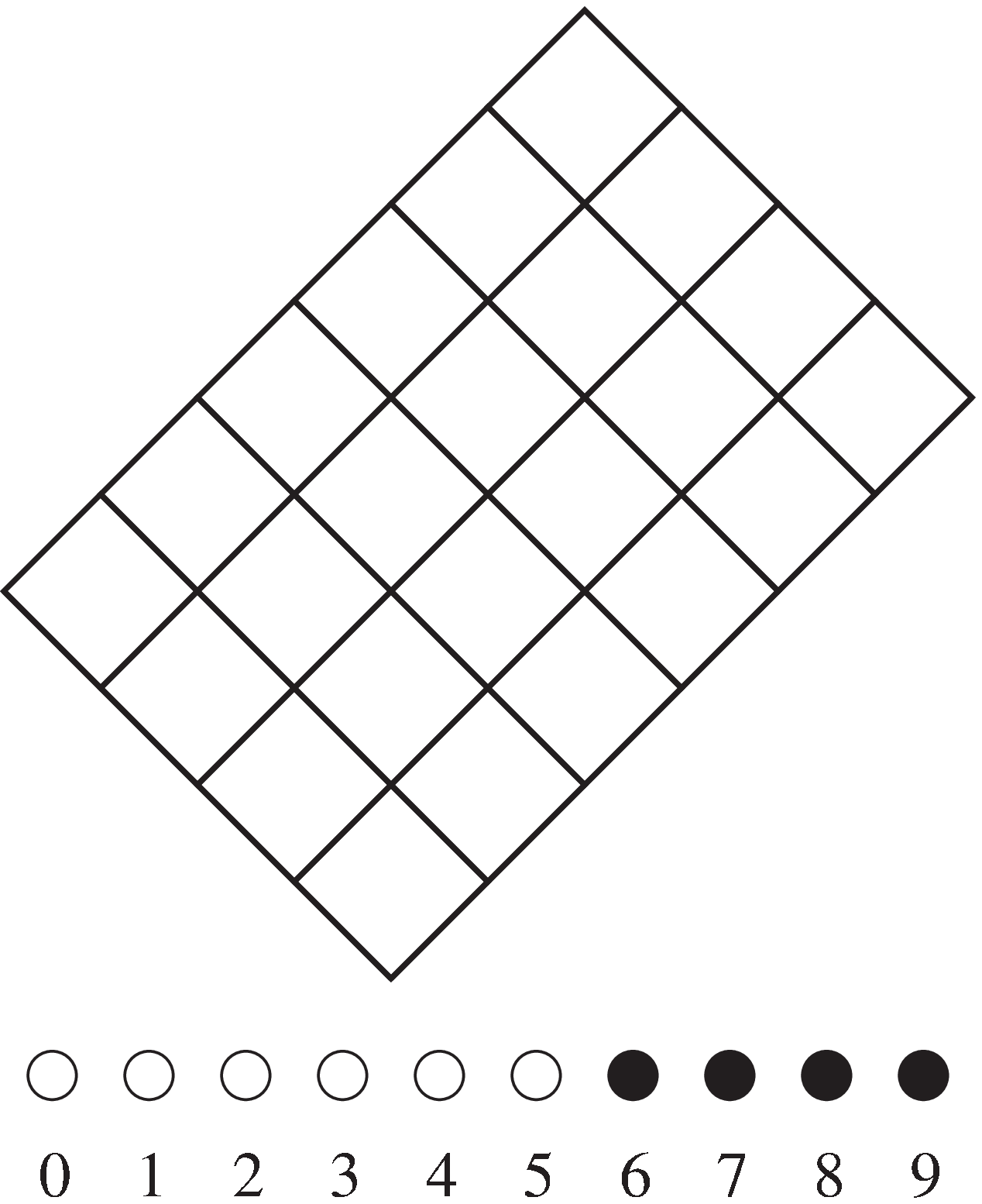}\\
(a) & (b) & (c)
\end{tabular}
\end{center}
\caption{An example of the correspondence between $\vec{\tilde{n}}$
and the restrictied Young diagram (partition) for $k=10$ and $N=4$.
The right-down and right-up edge of the boxes correspond to the black and white circle, respectively. 
(a) The empty box is for $\vec{\tilde{n}} = (3,2,1,0)$.
(b) The partition $\vec{\tilde{\nu}} = (6,3,2)$ is for $\vec{\tilde{n}} = (9,5,3,0)$.
(c) The full box is for $\vec{\tilde{n}} = (9,8,7,6)$.
The total number of the partitions is ${}_{10}C_4=210$,
which is the supersymmetric index of the $U(4)_{10}$ CS theory.
}
\label{Young diagrams}
\end{figure}

The $N$ ordered integers chosen from integers mod $k$ correspond to
the number of rows of the Young diagram within $N\times (k-N)$ boxes
by the identification
\be
\tilde{n}_a = \tilde{\nu}_a -a +(k-N),
\ee
where $\tilde{\nu}_a$ is the number of boxes of $a$-th row of the Young diagram
restricted within the $N\times (k-N)$ boxes. (See Figure~\ref{Young diagrams}.)
The supersymmetric index for $\N=2$ supersymmetric CS theory can be written as the partition function which is
normalized by the ground state partition function $\calZ_0$
\be
{\cal I}_{\N=2}(S^1\times \Sigma)
=
\frac{\calZ_\text{SYMCS}(S^1\times \Sigma)}{\calZ_0(S^1\times \Sigma)}
 = 
\sum_{R'}
(\dim_q R')^{\chi(\Sigma)},
\label{index}
\ee
where the sum is taken over the representation $R'$ associated with
the Young diagram within $N\times (k-N)$ boxes.

\begin{figure}[t]
\begin{center}
\begin{tabular}{ccc}
\includegraphics[scale=0.42]{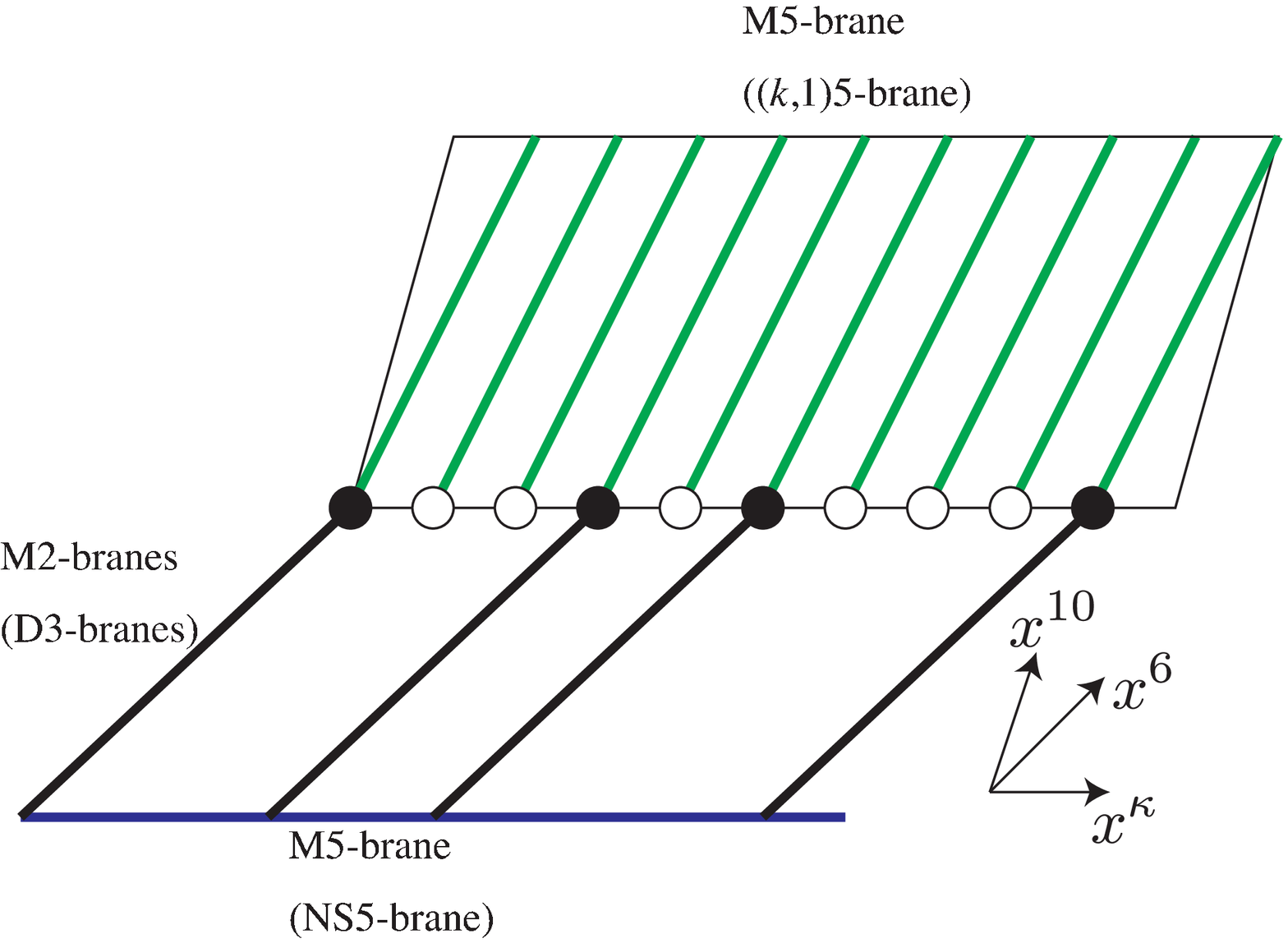}&&
\includegraphics[scale=0.36]{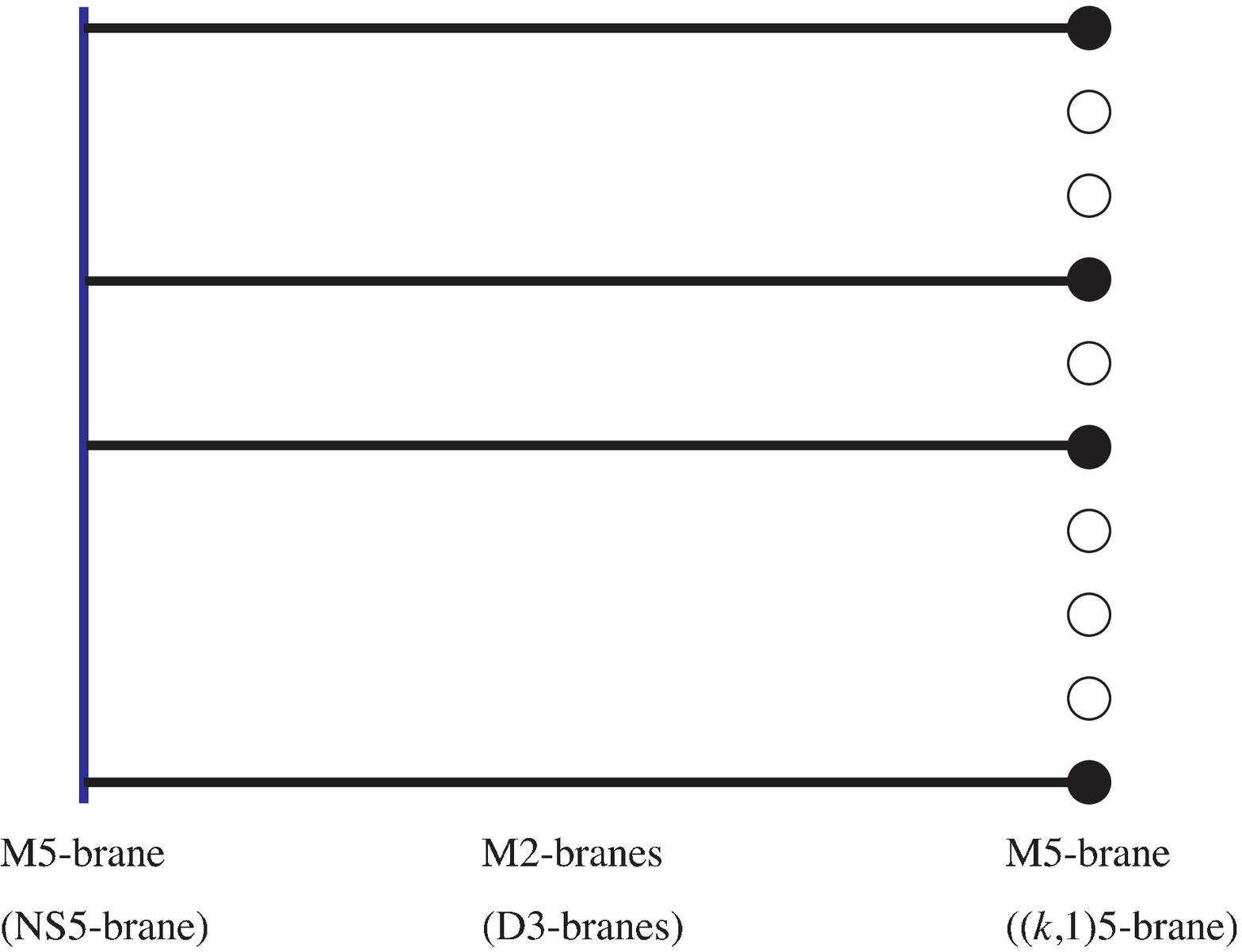}\\
(a) && (b)
\end{tabular}
\end{center}
\caption{
The supersymmetric YMCS theory can be realized by a configuration
with NS5-D3-$(k,1)$5 branes in Type IIB string theory.
The brane configuration is lifted up to a M5-M2-M5 system in M-theory on torus.
The positions of M2 branes,
which are interpreted by the vev of the Wilson loop,
are quantized along the fiber direction.
We depict the brane configuration by a diagram (b), which corresponds to
the restricted Young diagram in Figure \ref{Young diagrams}.
}
\label{M2-M5}
\end{figure}

In particular, for the $M=T^3$ case, namely $\chi(\Sigma)=0$, the index becomes
\be
{\cal I}_{\N=2}(T^3)
=
\sum_{R'} 1
 = \frac{k!}{N!(k-N)!},
\label{T3 index}
\ee
if $k\geq N$, since the number of the Young diagram within $N\times (k-N)$ boxes
is given by the binomial coefficient ${}_kC_{N}= \frac{k!}{N!(k-N)!}$
(the number of choices of $N$ integers within $k$).
The index (\ref{T3 index}) completely coincides with the value obtained in \cite{Witten:1999ds,Ohta:1999iv}.
(Incidentally, for $\N=1$ $U(N)$ supersymmetric CS theory,  the index is formally
 given by ${}_{k-\frac{N}{2}}C_{N}$ due to the quantum level shift $k\to k-\frac{N}{2}$.)
Thus we find that the choice of the Young diagram corresponds to
the choice of the supersymmetric vacua, which can be expressed by the brane configuration in
M-theory \cite{Ohta:1999iv}. (See also Figure \ref{M2-M5}.)
In this sense, the usual expression of the CS partition function via surgery,
where the partition function is normalized to be $Z(S^1\times S^2)=1$, 
relates to the proper vacuum which is called above the ``ground state''.\footnote{
The difference of the partition function between various vacua is just a phase.
So it is not suitable to call it the ground state since each vacuum is at zero energy and supersymmetric.}

The index (\ref{index}) is obviously invariant under the exchange $N \leftrightarrow (k-N)$.
This fact also reflects the level-rank duality (mirror symmetry)
between the three-dimensional supersymmetric YMCS theories of $U(N)$ at level $k$ and $U(k-N)$ 
at level $-k$
\cite{Kitao:1998mf}.

\subsection{The general $p$}

We now discuss the case of the general $p$ and $\mu=1$, where $M$ forms uniformly ``round'' shape
which includes the round $S^3$ for $p=1$.
We first decompose the integral region of $\phi_a$'s into the integer lattice following \cite{Blau:2006gh}
\be
\phi_a \equiv \tilde{\phi}_a + 2\pi n_a, \quad 0\leq \tilde{\phi}_a < 2\pi \text{ and } n_a \in \Z.
\ee
Then the partition function (\ref{phi integral of SYMCS}) becomes the integral over the compact regions of $\tilde{\phi}_a$
and the summation over $(\Z_p)^r$ and $\Z^r$
\begin{multline}
\calZ_\text{SYMCS}
 = \frac{1}{|W|}\sum_{\vec{m}\in (\Z_p)^r}
 \sum_{\vec{n}\in \Z^r}
\int_{0}^{2\pi} \prod_{a=1}^r \frac{d\tilde{\phi}_a}{2\pi}
\prod_{\alpha>0}
\left(
2\sin 
 \frac{\alpha(\tilde{\phi})}{2}
\right)^{\chi(\Sigma)}\\
\times
e^{i k \sum_{a=1}^r \left[\tilde{\phi}_a m_a+\frac{p}{4\pi} (\tilde{\phi}_a+2\pi n_a)^2\right]}.
\label{mu=1}
\end{multline}
Noting that
$e^{ikp\pi n_a^2} = e^{ikp\pi n_a}$
for the integer $k$, $p$ and $n_a$ and using agin the Poisson resummation formula
\begin{multline}
\frac{1}{(2\pi)^r}
\sum_{\vec{n}\in \Z^r}
\sum_{\vec{m}\in (\Z_p)^N}
\prod_{a=1}^r e^{ik\left[p (n_a+\pi)+m_a \right] \tilde{\phi}_a}\\
=
\begin{cases}
\frac{1}{k^r}\sum_{\vec{n}\in\Z^r}
\prod_{a=1}^r\delta\left(
\tilde{\phi}_a - \frac{2\pi n_a}{k}
\right) & \text{for even }kp\\
\frac{1}{(kp)^r}\sum_{\vec{n}\in(\Z+\frac{1}{2})^r}
\prod_{a=1}^r\delta\left(
\tilde{\phi}_a - \frac{2\pi n_a}{kp}
\right)e^{i\varphi(n_a)} & \text{for odd }kp
\end{cases},
\label{Poisson resum}
\end{multline}
where $e^{i\varphi(n_a)}\equiv \sum_{m\in \Z_p} e^{i\frac{2\pi}{p}n_a m_a}$.
Inserting (\ref{Poisson resum}) into (\ref{mu=1}), we can integrate $\tilde{\phi}_a$ explicitly, then
we get the partition function expressed as the $q$-deformed two-dimensional YM theory \cite{Blau:2006gh}
\be
\calZ^\text{even $kp$}_\text{SYMCS}
 = \frac{C}{k^r|W|}
 \sum_{\vec{n}\in \Z^r}
\prod_{\alpha>0}
[\alpha(n)]_q^{\chi(\Sigma)}
q^{\frac{1}{2}p \sum_{a=1}^r n_a^2},
\ee
if $kp$ is even, where $q=e^{\frac{2\pi i}{k}}$.
In particular, if we choose $G=U(N)$ and $M=S^3$, namely $p=1$ and $\chi(\Sigma)=2$, we obtain
the partition function on $S^3$
\be
\calZ_\text{SYMCS}(S^3)
 = \frac{C}{k^N|W|}
 \sum_{\vec{n}}
\prod_{1\leq a< b \leq N}
[n_a-n_b]_q^{2}
q^{\frac{1}{2}\sum_{a=1}^N n_a^2},
\ee
where the summations are taken over $(\Z_k)^N$ for even $k$ and $(\Z_k+\frac{1}{2})^N$ for odd $k$.

If $p>1$ in general and $kp$ is odd,
the expression of the partition function becomes more complicated due to the phase factors.
A similar dependence on $k$ whether even or odd
 is reported in the exact evaluation of the localized integral \cite{Okuyama:2011su}.

\subsection{The Wilson loop}

As explained above, the supersymmetric Wilson loop operator is $Q$-closed but not $Q$-exact.
So we can also evaluate the vev of the Wilson loops without violating the localization structure
in the supersymmetric YMCS theory.

At the localization fixed point, the Wilson loop takes the value
\bea
\left. W(C) \right|_\text{fixed points} &=&\Tr_R e^{i \phi}\nn\\
&=&
\frac{\sum_{w\in W} (-1)^w \e(w)e^{i \, w(\rho + \Lambda_R)\cdot \phi}}
{\sum_{w\in W} (-1)^w \e(w)e^{i  \, w(\rho)\cdot \phi}}\nn\\
&\equiv& W_R(\phi)
,
\eea
where $w$ is the element of the Weyl group $W$ and $\Lambda_R$ 
is the highest weight (Young diagram) corresponding to the representation $R$.
Furthermore, as we have seen,
using the Poisson resummation and integrating $\phi$ explicitly
in the $p\to \infty$ limit, $\phi$'s are fixed at the
discrete integer set $\phi_a=2\pi n_a / k$.
The vev of the Wilson loop can be evaluated by
\be
\left\langle W(C) \right\rangle
=
\frac{\calZ_\text{SYMCS}^{-1}}{k^r|W|}\sum_{\vec{n}\in \Z^r}
\prod_{\alpha>0}
\left(
2\sin 
\frac{\pi}{k} \alpha(n)
\right)^{\chi(\Sigma)}
W_R\left(\frac{2\pi \vec{n}}{k}\right)
e^{i \frac{\pi}{k}\mu \sum_{a=1}^r n_a^2 }.
\ee
If we restrict the sum at the ground state $n_a=\rho_a$ only,
then it recovers the (unknot) Wilson loop in the bosonic CS theory on $S^3$ obtained from the surgery
\be
\left.\left\langle W(C) \right\rangle\right|_\text{ground state}
=
W_R\left(\frac{2\pi \vec{\rho}}{k}\right)
= \prod_{\alpha>0}
\frac{\sin \frac{\pi}{k} \alpha(\Lambda_R+\rho)}{\sin \frac{\pi}{k} \alpha(\rho)}
=\dim_q R,
\ee
with $q=e^{\frac{2\pi i}{k}}$.

\section{Including Matters}

\subsection{Localization for chiral superfield}

So far, we have investigated the supersymmetric YMCS theory without matter multiplets.
In this section, we would like to include the matters following the formulation of the
equivariant cohomology introduced above.

First of all, we start with the $\N=2$ supersymmetric transformations 
of the chiral superfield in three-dimensions as well as the vector (gauge) multiplet.
The chiral superfield consists of the complex scalar $X$, the chiral fermion $\psi$
and the auxiliary complex scalar field $F_c$. 
We do not consider the mass for the chiral superfield for a while.
We will introduce the mass of the field later.

As well as the vector multiplet,
we obtain the BRST transformations from the $\N=2$  supersymmetric transformations
for the chiral superfields
(see Appendix B)
\be
\begin{array}{ll}
Q X = \psi, & Q \psi = -i(\D_\kappa X -i \sigma \cdot X),\\
Q Y_\zb =-i(\D_\kappa \chi_\zb -i \sigma\cdot \chi_\zb), & Q \chi_\zb =Y_\zb.
\label{BRST transformation chiral}
\end{array}
\ee
Note that $Y_\zb$, $\chi_\zb$ and their hermite conjugate
coming from the anti-chiral superfield
act as complex vector fields on $\Sigma$.
So introducing the (0,1)-form on $\Sigma$ by
$Y_c \equiv Y_\zb d\zb$ for the bosonic fields and
$\chi_c \equiv \chi_\zb d\zb$, then
the BRST transfomations can be written  by a simpler form
\be
\begin{array}{ll}
Q X = \psi, & Q \psi = -i(\cL_V X -\delta_\Phi X),\\
Q Y_c =-i(\cL_V \chi_c-\delta_\Phi \chi_c), & Q \chi_c =Y_c,
\end{array}
\ee
where $\delta_\Phi X= i\Phi \cdot X$ is the gauge transformation
with respect to the gauge parameter $\Phi$
and
the coupling with the vector multiplets $\Phi \cdot X$ 
depends on the representation of the chiral superfield.
Thus we have constructed the BRST symmetry, which satisfies obviously
\be
Q^2 = -i(\cL_V -\delta_\Phi).
\ee

Therefore we can argue the localization for the chiral superfields by using the equivariant cohomology of $Q$
in the same way as the vector multiplet.
The $Q$-exact action, which is needed for the localization,
is nothing but the matter part of the supersymmetric YM action.
Introducing the vector notation of the matter fields $\vec{\B}_m=(X,Y_c)$ for the bosons and 
$\vec{\F}_m=(\psi,\chi_c)$ for the fermions, the action can be written by
\be
S_\text{matter}
=\frac{1}{2}
Q\int_M
\Tr\left[
\vec{\F}_m \wedge \star \overline{Q\vec{\F}_m}
-2i \chi_c \wedge \star \mu_c^\dag
-2i \chi_c^\dag \wedge \star \mu_c
\right],
\ee
where $\mu_c\equiv \pi^* d_{A} X+i\frac{\del W(X)}{\del X}\kappa$ is the moment map which gives the F-term constraint
with the superpotential $W(X)$.
The bosonic part of $S_\text{matter}$ becomes quadratic and positive definite in $Q\vec{\F}_m$ and $\mu_c$
\be
\left. S_\text{matter} \right|_\text{boson}
=\frac{1}{2}
\int_M
\Tr \left[
Q \vec{\F}_m \wedge \star \overline{Q\vec{\F}_m} + 2\mu_c \wedge \star \mu_c^\dag
\right],
\ee
after integrating out the auxiliary field $Y_c$.
So we can conclude that the path integral for the matter fields is localized at
the BRST fixed point $Q\vec{\F}_m=0$ and $\mu_c=0$
by using the coupling independence of the $Q$-exact action.
Note that the F-term constraint $\mu_c=0$ is strict on the localization
while the D-term constraint admits the higher critical points $\mu_r\neq 0$,
since the F-term constraint does not relate to the complexification of $\Phi$.
So we always have to take into account the F-term constraints
when we are considering the localization fixed points.

The contribution from the chiral matter fields to the 1-loop determinants can be
derived in the similar way as the vector multiplet case.
It is exactly given by 
\bea
e^{-S_\text{matter}} 
&=&\sqrt{
\frac{
\Det \frac{\delta (Q\B_m^I)}{\delta \F_m^J}}
{\Det \frac{\delta (Q\F_m^I)}{\delta \B_m^J}}}
\delta(Q\vec{\F}_m)\delta(\vec{\F}_m)\nn\\
&=&
\frac{\Det_{\chi_c}(-i\cL_V-\rho(\phi))}
{\Det_{X}(-i\cL_V-\rho(\phi))}
\delta(Q\vec{\F}_m)\delta(\vec{\F}_m),
\label{matter 1-loop}
\eea
where $\rho(\phi)$ is the weights of  the representation $R$ of $X$,
that is, $\rho(\phi)=\alpha(\phi)$ in the adjoint representation 
and $\rho(\phi)=\phi_a$ in the fundamental representation, for example.
We here note that the fields in the chiral superfields are complex valued
and the determinant admits some phases because of the absence of the absolute value.


Now let us introduce the mass for the matter fields.
There are two possibilities to give the mass to the chiral superfields:

One comes from the F-term. The quadratic term in the superpotential induces the explicit mass term for $X$.
As we mentioned above, the F-term constraint is important to solve the fixed point equation. The fixed point
equation sometimes removes the zero modes of the matter fields at the low energy.
We give an example of this case in the next subsection.

The other is introduced 
as the eigenvalue of the Lie derivative $\cL_V$. In fact,
if we assume that the matter fields has a twisted boundary condition
\be
\Psi(z,\zb,\theta+2\pi\ell) = \Psi(z,\zb,\theta) e^{2\pi i m},
\ee
along the fiber direction,
the boundary condition induces the mass for $X$ by the so-called Scherk-Schwarz mechanism
\cite{Scherk:1978ta}.
This also can be understood as the mass term induced by the $\Omega$-background \cite{Nekrasov:2003rj}
from the point of view of two-dimensional theory on $\Sigma$. In this sense, the eigenvalues of $\cL_V$
for the matter fields are modified to
$i \left(-\frac{n}{\ell} + \frac{m}{\ell} \right)$ for $n\in\Z$.
Here $m$ is the mass of the matter fields measured in the unit of the Kaluza-Klein mass scale
of the $S^1$ fiber.

In conformal field theory on $M$, the mass is determined by the dimension of fields.
If we assign the dimension $\Delta$ to the lowest component $X$, then
$\psi$, $\chi_c$ and $Y_c$ also have the same dimension $\Delta$,
because of the twisting of our formulation. (See Appendix B.)
In the canonical assignments, we should set $\Delta=1/2$.
We must impose the boundary conditions of  these fields to get the correct mass with the dimensions.
It is interesting that all fields in the matter multiplet have the fermionic twisted boundary condition
along the $S^1$ fiber
\be
\Psi(z,\zb,\theta+2\pi\ell) = -\Psi(z,\zb,\theta),
\label{fermionic boundary condition}
\ee
for the conformal matter with the canonical dimension  $\Delta=1/2$ on $M$.

Once the superpotential vanishes and
the eigenvalues of $\cL_V$ are given by the conformal dimension,
we can evaluate the 1-loop determinant (\ref{matter 1-loop})
explicitly.
If we expand the fields by the eigenmodes of $\cL_V$
\be
X(z,\bar{z},\theta) = \sum_{n\in\Z} X_n(z,\bar{z})e^{-i (n-\Delta)\theta/\ell},
\quad
\chi_c(z,\bar{z},\theta) = \sum_{n\in\Z} \chi_{c,n}(z,\bar{z})e^{-i (n-\Delta)\theta/\ell}.
\ee
%
%
%
Noting that 
$X_{n}$ and $\chi_{c,n}$ have the same eigenvalue and
the 0-form and 1-form sections
of the line bundles $\Op(-pn+\lfloor p\Delta \rfloor)$ over $\Sigma$, respectively, where
$\lfloor x \rfloor \equiv \max\{n\in\Z|n\leq x\}$ is the floor function,
the difference of the number of these modes on $\Sigma$
between $X_{n}$ and $\chi_{c,n}$ can be obtained from
the Hirzebruch-Riemann-Roch theorem
\begin{multline}
\dim \Omega^0(\Sigma,\Op(-pn+\lfloor p\Delta \rfloor)\otimes V_\rho)
-\dim \Omega^1(\Sigma,\Op(-pn+\lfloor p\Delta \rfloor)\otimes V_\rho)\\
=\frac{1}{2}\chi(\Sigma)-pn+\lfloor p\Delta \rfloor +\rho(m),
\end{multline}
where $\rho(m)$ is the contribution from the first Chern class of the
$U(1)$ gauge fields in the matter representation $R$,
as well as the vector multiplet. 
Since there are infinitely many pairs of $X_{n}$ and $\chi_{c,n}$ as the fields on $\Sigma$,
 the contribution from the matter to the 1-loop determinant is given by
\be
\prod_{n\in \Z}
\frac{\Det_{\chi_{c,n}}(-i\cL_V-\rho(\phi))}
{\Det_{X_{n}}(-i\cL_V-\rho(\phi))}
=\prod_{\rho\in R} \prod_{n\in\Z}
\left(-\frac{n-\Delta}{\ell}-\rho(\phi)\right)^{-\frac{1}{2}\chi(\Sigma)+pn - \lfloor p\Delta \rfloor-\rho(m)}.
\label{matter determinant}
\ee
In particular, if we consider the case of
$M=S^3$ ($\chi(\Sigma)=2$ and
$p=1$) and assume $\Delta<1$, then
the determinant becomes
\be
\prod_{\rho\in R} \prod_{n\in\Z}
\left(-\frac{n-\Delta}{\ell}-\rho(\phi)\right)^{n-1}
=
\prod_{\rho\in R} \prod_{n=1}^{\infty}
\left(-
\frac{n+1-\Delta+\ell\rho(\phi)}{n-1+\Delta-\ell\rho(\phi)}
\right)^n,
\ee
which agrees with the results in \cite{Kapustin:2009kz,Hama:2010av,Marino:2011nm} up to a phase factor.

The determinant (\ref{matter determinant}) contains the phase in general,
but we are interested in the matters only in the self-conjugate representation
in the present paper.
So we now concentrate on the absolute value of the
determinant, which becomes
\bea
\prod_{\rho\in R} \prod_{n\in\Z}
\left|\frac{n-\frac{1}{2}}{\ell}+\rho(\phi)\right|^{-\frac{1}{2}\chi(\Sigma)+pn-\lfloor p/2 \rfloor-\rho(m)}
&=&
\prod_{\rho\in R} \prod_{n=1}^{\infty}
\left|
\frac{\left(n-\frac{1}{2}\right)^2}{\ell^2}-\rho(\phi)^2
\right|^{-\frac{1}{2}\chi(\Sigma)-\lfloor p/2 \rfloor-\rho(m)}\nn\\
&&
\qquad\qquad
\times
\left|
\frac{n-\frac{1}{2}+\ell \rho(\phi)}
{n+\frac{1}{2}-\ell \rho(\phi)}
\right|^{pn}\nn\\
&=&\prod_{\rho\in R}
\left|
2\ell \cos\pi \ell \rho(\phi)
\right|^{-\frac{1}{2}\chi(\Sigma)-\lfloor p/2 \rfloor-\rho(m)}\nn\\
&&\qquad\qquad
\times
\prod_{n=1}^\infty
\left|
\frac{n-\frac{1}{2}+\ell \rho(\phi)}
{n+\frac{1}{2}-\ell \rho(\phi)}
\right|^{pn}
,
\eea
when the matter multiplet has the canonical dimension $\Delta=1/2$,
where we have used the infinite product expression of the cosine-function and the zeta function regularizations.
If the matter field is in the adjoint or  self-conjugate 
(hypermultiplet) representation,
the dependence of the phase and $\rho(m)$
disappears and the determinant simplifies to
\be
\ell^{-\chi(\Sigma)\dim G}\prod_{\alpha>0}\left(
2 \cos\pi \ell \alpha(\phi)
\right)^{-\chi(\Sigma)+\varepsilon(p)},
\ee
for the adjoint representation and
\be
 \ell^{-\chi(\Sigma)r}\prod_{a=1}^r
\left(2\cos\pi \ell \phi_a
\right)^{-\chi(\Sigma)+\varepsilon(p)},
\ee
for the self-conjugate fundamental representation $\Box \oplus \overline{\Box}$,
where $\varepsilon(p)\equiv p -2\lfloor p/2 \rfloor =0 \text{ or } 1$ such that $p=\varepsilon(p) \text{ mod } 2$.
The appearance of the cosine function $\cos x\sim 1+ e^{ix}$ might be  related to the fermionic nature
(\ref{fermionic boundary condition}) of the matter fields
with the canonical dimension.
These formulae agree with the results in \cite{Kapustin:2009kz} for 
$M=S^3$.

\subsection{$\N=3$ YMCS theory}

We here comment on the case that
 the $\N=2$ supersymmetric YMCS theory with a single massive chiral superfield
in the adjoint representation, whose mass is given by the superpotential.
The on-shell fields in the vector multiplet of the YMCS theory have
the topologically induced mass of $kg^2/4\pi$, where $g$ is the gauge coupling constant.
If we tune the mass of the adjoint chiral superfield to $kg^2/4\pi$ which is the same mass as the vector multiplet,
the $\N=2$ supersymmetry enhances to $\N=3$
since $SO(3)$ R-symmetry now acts on
the triplet of the adjoint scalars in the $\N=2$ vector multiplet
and chiral superfield .
This bare complex mass is given by the superpotential $W= \frac{k}{8\pi}\Tr X^2$.
The brane realization of this system is discussed in \cite{Kitao:1998mf}.

However this is essentially in the UV picture. In the IR limit, all massive fields compared with
the mass scale $kg^2/4\pi$ in the supersymmetric YMCS theory
are integrated out and flows to the $\N=3$ supersymmetric CS theory with the conformal symmetry.
At the conformal fixed points, 
the adjoint matter $X$ decouples and the residual matter contents reduce to
the $\N=2$ vector multiplet only.
In the localization language, this is the consequence of the fixed point equation $X=0$ from
the F-term constraint $\mu_c=0$.

Thus, in the low energy limit, we obtain the partition function of the $\N=3$ 
supersymmetric YMCS theory which is the same as the $\N=2$ one.
Then, of course, the supersymmetric index of this system on $T^3$ ($\chi(\Sigma)=0$)
is the same as the index for $\N=2$ theory
\be
{\cal I}_{\N=3}(T^3) = \frac{k!}{N!(k-N)!}.
\ee
This agrees with the observation in \cite{Ohta:1999iv}.

\subsection{ABJM theory}

We now apply the localization procedure to the ABJM theory \cite{Aharony:2008ug}
or its generalization \cite{Aharony:2008gk}.

The theory has the product of the gauge groups $U(N+l)\times U(N)$
and there are the CS couplings with the opposite level $k$ and $-k$.
We denote this by the symbol $G=U(N+l)_k\times U(N)_{-k}$
and assume $l\geq 0$ in the following.
The ABJM theory is the case of $l=0$ and we call the ABJ theory for the general $l$.
In the UV  picture, the YMCS theory contains a set of the vector multiplets
$(A,\sigma,Y_r;\lambda,\eta,\chi_r)$ and 
$(\tilde{A},\tilde{\sigma},\tilde{Y}_r;\tilde{\lambda},\tilde{\eta},\tilde{\chi}_r)$,
the chiral superfield in the adjoint representation
$(X,Y_c;\psi,\chi_c)$ and 
$(\tilde{X},\tilde{Y}_c;\tilde{\psi},\tilde{\chi}_c)$,
four chiral superfields
$(Z_i,Y^{Z_i}_c;\psi^{Z_i},\chi^{Z_i}_c)$ in the bifundamental and 
and $(\tilde{Z}_i,Y^{\tilde{Z}_i}_c;\psi^{\tilde{Z}_i},\chi^{\tilde{Z}_i}_c)$
in the anti-bifundamental representation
 for $i=1,2$.

The BRST transformations are
\be
\begin{array}{ll}
Q A = \lambda, & Q \lambda = -i(\cL_V A -d_A\Phi),\\
Q \tilde{A} = \tilde{\lambda}, & Q \tilde{\lambda} = -i(\cL_V \tilde{A} -d_{\tilde{A}}\tilde{\Phi}),\\
Q\Phi =0, & Q\Phit =0,  \\
Q\Phib = 2\eta, & Q\eta = -i(\cL_V \Phib -i[\Phi,\Phib]), \\
Q\bar{\tilde{\Phi}} = 2\tilde{\eta}, & Q\bar{\eta} = -i(\cL_V \bar{\tilde{\Phi}} -i[\Phi,\bar{\tilde{\Phi}}]), \\
Q Y_r =-i(\cL_V \chi_r- i[\Phi,\chi_r]), & Q \chi_r =Y_r,\\
Q \tilde{Y}_r =-i(\cL_V \tilde{\chi}_r- i[\tilde{\Phi},\tilde{\chi}_r]), & Q \tilde{\chi}_r =\tilde{Y}_r,
\end{array}
\ee
for the vector multiplets, where $\Phi=\iota_V A +i \sigma$ and $\tilde{\Phi}=\iota_V \tilde{A} +i \tilde{\sigma}$,
\be
\begin{array}{ll}
Q X = \psi, & Q \psi = -i(\cL_V X -i [\Phi, X]),\\
Q \Xt = \psit, & Q \psit = -i(\cL_V \Xt -i [\Phit,\Xt]),\\
Q Y_c =-i(\cL_V \chi_c-i [\Phi, \chi_c]), & Q \chi_c =Y_c,\\
Q \tilde{Y}_c =-i(\cL_V \tilde{\chi}_c-i[\Phit,\tilde{\chi}_c]), & Q \tilde{\chi}_c =\tilde{Y}_c,
\end{array}
\ee
for the chiral superfields in the adjoint representation, and
\be
\begin{array}{ll}
Q Z_i = \psi^{Z_i}, & Q \psi^{Z_i} = -i(\cL_V Z_i -i (\Phi Z_i-Z_i\Phit)),\\
Q \tilde{Z}_i = \psi^{\tilde{Z}_i}, & Q \psi^{\tilde{Z}_i} = -i(\cL_V \tilde{Z}_i -i (\Phit \tilde{Z}_i - \tilde{Z}_i\Phi)),\\
Q Y_c^{Z_i} =-i(\cL_V \chi_c^{Z_i} -i (\Phi \chi_c^{Z_i}-\chi_c^{Z_i} \Phit)), & Q \chi_c^{Z_i} =Y_c^{Z_i},\\
Q Y_c^{\tilde{Z}_i} =-i(\cL_V \chi_c^{\tilde{Z}_i} -i(\Phit \chi_c^{\tilde{Z}_i}-\chi_c^{\tilde{Z}_i} \Phi)), & Q \chi_c^{\tilde{Z}_i} =Y_c^{\tilde{Z}_i},
\end{array}
\ee
for the bifundamental and anti-bifundamental matters.
The moment maps (D-term and F-term constraints) corresponding to the auxiliary fields are given by
\be
\begin{array}{lcl}
\mu_r &=& \frac{\kappa \wedge F_\Sigma}{\kdk}
+[X,X^\dag]
+\sum_{i=1,2}(|Z_i|^2-|\tilde{Z}_i|^2)
, \\
\tilde{\mu}_r &=& \frac{\kappa \wedge \tilde{F}_\Sigma}{\kdk}
+[\Xt, \Xt^\dag]
-\sum_{i=1,2}(|Z_i|^2-|\tilde{Z}_i|^2)
, \\
\mu_c &=& \pi^*d_\Sigma X-i[A,X] +i \kappa \frac{\del W}{\del X}, \\
\tilde{\mu}_c &=& \pi^*d_\Sigma \Xt - i[A,\Xt]+ i \kappa \frac{\del W}{\del \Xt},\\
\mu_c^{Z_i} &=& \pi^*d_\Sigma Z_i - iAZ_i+iZ_i \tilde{A} +i \kappa \frac{\del W}{\del Z_i}, \\
\mu_c^{\tilde{Z}_i} &=& \pi^*d_\Sigma\tilde{Z}_i-i\tilde{A}\tilde{Z}_i+i\tilde{Z}_i A+ i\kappa \frac{\del W}{\del \tilde{Z}_i},
\end{array}
\ee
where
\be
W = \Tr\left[
\frac{k}{8\pi}X^2 - \frac{k}{8\pi}\Xt^2
+\sum_{i=1,2}\left(
Z_i X \tilde{Z}_i + \tilde{Z}_i \Xt Z_i
\right)
\right],
\ee
is the superpotential of the UV lagrangian for the $\N=3$ supersymmetric YMCS theory
with the ABJM (ABJ) matter contents. 

We can construct a $Q$-exact YM action from the above BRST transformations and
the moment maps as well as we have discussed so far.
And also, 1-loop determinants are obtained from the field derivatives of the
BRST transformations.
However all fields do not contribute to the determinants because the massive fields
are integrated out and disappear in the low energy limit.


The adjoint scalars $X$ and $\Xt$ are massive by the superpotential, so we can integrate out these fields.
The superpotential reduces to
\be
W = \frac{4\pi}{k}\Tr
\left[
Z_1 \tilde{Z}_1 Z_2 \tilde{Z}_2
-Z_1 \tilde{Z}_2 Z_2 \tilde{Z}_1
\right],
\ee
which leads to the enhancement of the supersymmetry at the conformal point.
In this low energy limit, the adjoint scalars $X$ and $\Xt$ disappears.
The partition function is obtained from the 1-loop determinants of the vector multiplets
and the (anti-)bifundamental mattes. The result is 
\begin{multline}
\calZ_{\text{ABJ}}^{U(N+l)_k\times U(N)_{-k}}\\
 = \frac{1}{(N+l)!N!}\sum_{\vec{m},\vec{\tilde{m}}}
\int_{-\infty}^\infty \prod_{a=1}^{N+l} \frac{d\phi_a}{2\pi}
\prod_{c=1}^{N} \frac{d\tilde{\phi}_c}{2\pi}
\frac{
\left(
\prod_{a<b}2\sin 
\frac{\phi_a-\phi_b}{2}\prod_{c<d} 2 \sin 
\frac{\tilde{\phi}_c-\tilde{\phi}_d}{2}
\right)^{\chi(\Sigma)}}
{\prod_{a, c}\left( 2\cos
\frac{\phi_a-\tilde{\phi}_c}{2}
\right)^{2(\chi(\Sigma)-\varepsilon(p))}}
\\
\times
e^{ik \left[
\sum_{a=1}^{N+l} (\phi_a m_a
+ \frac{p\mu}{4\pi}\phi_a^2)
-\sum_{a=1}^{N} (\tilde{\phi}_a \tilde{m}_a
+\frac{p\mu}{4\pi} \tilde{\phi}_a^2)
\right]},
\label{ABJ}
\end{multline}
where we have used the fact that the bifundamental matters, which are
in the self-conjugate representation, have the canonical dimension $\Delta=1/2$.
We ignore the overall scale factor in the following.
The partition function (\ref{ABJ}) agrees with the result for $S^3$ \cite{Kapustin:2009kz,Marino:2011nm},
which is the case of $p=\mu=1$ and $\chi(\Sigma)=2$ of ours.

\begin{figure}[t]
\begin{center}
\includegraphics[scale=0.45]{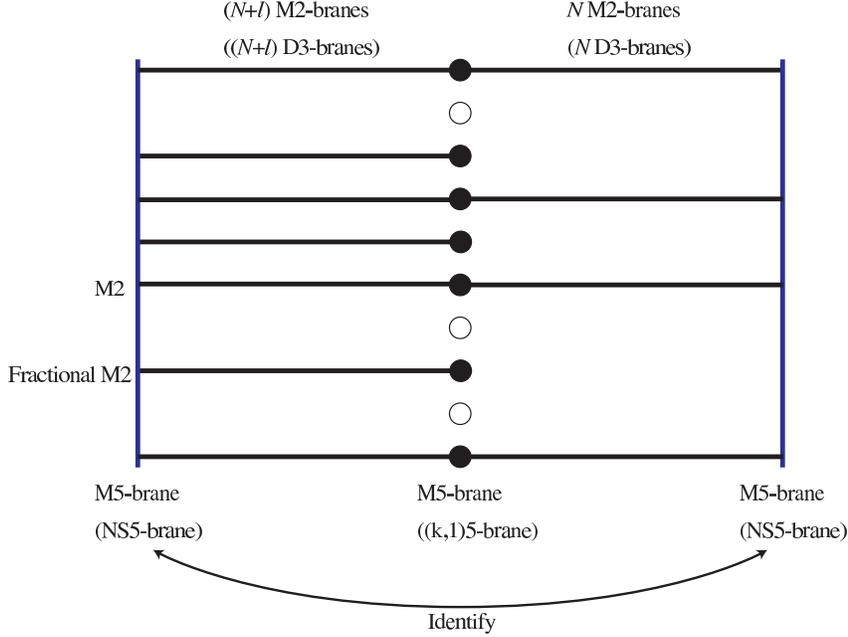}
\end{center}
\caption{
A vacuum configuration of the ABJ theory with the gauge group $U(N+l)_k \times U(N)_{-k}$.
The branes are placed on the compact circle along the $x^6$-direction. So
left and right M5(NS5)-branes are identified on the circle.
If positions of fractional M2-branes coincide with each other, they become a single M2-brane
wrapping around the $x^6$-direction and are decoupled from the M5-brane system
(or can be freely removed from the tip of the $\C^4/\Z_k$ orbifold in the dual M-theory picture).
Thus the partition function of the ABJ theory
is factorized into the fractional M2-brane sector (CS theory) and ordinary
(non-fractional) M2-brane sector (ABJM theory).}
\label{ABJ diagram}
\end{figure}

Let us first consider the supersymmetric index for this system.
Setting $p=0$ and using the Poisson resummation formula,
the partition function on $M=S^1\times \Sigma$ becomes
\begin{multline}
\calZ_{\text{ABJ}}^{U(N+l)_k\times U(N)_{-k}}(S^1\times \Sigma)\\
 = \frac{1}{(N+l)!N!}\sum_{\vec{n},\vec{\tilde{n}}}
\left(
\frac{
\prod_{a<b}2\sin 
\frac{\pi}{k}(n_a-n_b)
\prod_{c<d} 2\sin 
\frac{\pi}{k}(\tilde{n}_c-\tilde{n}_d)
}
{\prod_{a, c} (2\cos
\frac{\pi}{k}(n_a-\tilde{n}_c))^2}
\right)^{\chi(\Sigma)},
\label{ABJ index}
\end{multline}
where $\vec{n}\in (\Z_k)^{N+l}$ and $\vec{\tilde{n}}\in (\Z_k)^N$.
This partition function includes various phases (vacuum configurations),
but we can easily classify them graphically by using the brane configuration
(the M2-M5 system on M-theory torus), see Figure \ref{ABJ diagram}.
If some $n_a$ and $\tilde{n}_b$ coincide with each other,
these M2 branes are decoupled from the bound state with the M5-brane,
since the connected fractional M2's are promoted a single M2.
The maximal number of the connected fractional M2 is $N$ for the ABJ theory.
In this maximal phase, the partition function (\ref{ABJ index}) is factorized into
\be
\calZ_\text{ABJ}^{U(N+l)_k\times U(N)_{-k}}(S^1\times \Sigma)
= \calZ_{\N=2}^{U(l)_k}(S^1\times \Sigma) \times \calZ_\text{ABJM}^{U(N)_k\times U(N)_{-k}}
(S^1\times \Sigma),
\label{partition function product}
\ee
where $\calZ_{\N=2}^{U(l)_k}$ is the same partition function as the pure $\N=2$ (or equivalently $\N=3$) theory
with the gauge group $G=U(l)_k$ on $S^1\times \Sigma$ and 
\be
\calZ_{\text{ABJM}}^{U(N)_k\times U(N)_{-k}}(S^1\times \Sigma)
 = \frac{1}{N!}\sum_{\vec{n}\in (\Z_k)^N}
\prod_{1\leq a<b \leq N}
\left(
\frac{
\tan 
\frac{\pi}{k} (n_a-n_b)}
{2^{2N}\cos \frac{\pi}{k} (n_a-n_b)}
\right)^{2\chi(\Sigma)},
\label{ABJM}
\ee
since $n_a=\tilde{n}_c$ for the connected fractional M2.
This factorization also can be understood from the brane configuration in Figure~\ref{ABJ diagram}.

If we now consider the case of $\chi(\Sigma)=0$, namely $M=T^3$, we get the supersymmetric index as the number
of terms (configurations) in the partition function (\ref{partition function product}). Then we find
\be
{\cal I}_\text{ABJ}^{U(N+l)_k\times U(N)_{-k}}
= {\cal I}_{\N=2}^{U(l)_k} \times {\cal I}_\text{ABJM}^{U(N)_k\times U(N)_{-k}},
\ee
where
\bea
&&{\cal I}_{\N=2}^{U(l)_k}={}_kC_l = \frac{k!}{l!(k-l)!},\\
&& {\cal I}_\text{ABJM}^{U(N)_k\times U(N)_{-k}}={}_kC_N = \frac{k!}{N!(k-N)!}.
\eea
This index is also obvious from the brane configuration.
Using the mirror symmetry between $U(l)_k$ and $U(k-l)_{-k}$ theories,
which means ${\cal I}_{\N=2}^{U(l)_k} = {\cal I}_{\N=2}^{U(k-l)_{-k}}$,
we can immediately see the equivalence of the supersymmetric index
\be
{\cal I}_\text{ABJ}^{U(N+l)_k\times U(N)_{-k}}={\cal I}_\text{ABJ}^{U(N)_k\times U(N+k-l)_{-k}}.
\ee
This is a proof of the equivalence suggested in \cite{Aharony:2008gk} at the supersymmetric index level.

Let us  pay attention to only the ABJM partition function, which is the case of $l=0$ in (\ref{ABJ}),
 in the following discussions.
Assuming $p \to \infty$ with fixed $p\mu$ or $\mu=1$ with even $kp$,
the partition function reduces to the discrete sum
over $(\vec{n},\vec{\tilde{n}})\in (\Z_k)^{2N}$ as discussed in the pure CS theory
\begin{multline}
\calZ_{\text{ABJM}}^{U(N)_k\times U(N)_{-k}}
 = \frac{1}{(N!)^2}\sum_{\vec{n},\vec{\tilde{n}}}
\frac{
\prod_{a<b}
\left(2\sin \frac{\pi}{k} (n_a-n_b)\right)^{\chi(\Sigma)}
\left(2 \sin \frac{\pi}{k} (\tilde{n}_a-\tilde{n}_b)\right)^{\chi(\Sigma)}}
{\prod_{a, b}\left( 2\cos
\frac{\pi}{k}(n_a-\tilde{n}_b)
\right)^{2(\chi(\Sigma)-\varepsilon(p))}}
\\
\times
e^{i\frac{\pi}{k}p\mu 
\sum_{a=1}^{N}
(n_a^2
-
\tilde{n}_a^2
)},
\label{general ABJM}
\end{multline}
which contains the fractional M2 branes in general.
The coincide integers $n_a=\tilde{n}_b$ stands for the connected fractional M2's, which
can decouple from the CS coupling (remove from the tip of $\C^4/\Z_k$)
since the corresponding quadratic term from the CS coupling vanishes.
In particular, if all of the fractional M2 branes connect with each other, that is
$\tilde{n}_a=n_a$ with a fixed order, then we obtain the partition function for the non-fractional $N$ M2 branes
\be
\calZ_N^\text{M2}(M)
 = \frac{1}{2^{2N(\chi(\Sigma)-\varepsilon(p))}N!}\sum_{\vec{n}\in (\Z_k)^N}
\prod_{a<b}
\frac{
\left(2\sin \frac{\pi}{k} (n_a-n_b)\right)^{2\chi(\Sigma)}
}
{\left( 2\cos
\frac{\pi}{k}(n_a-n_b)
\right)^{4(\chi(\Sigma)-\varepsilon(p))}}.
\ee
It it interesting to discuss the large $N$ behavior of the above partition function written in terms of
the discrete sum, but we leave it for future work.

\section{Conclusion and Discussion}

In this paper, we have investigated the supersymmetric YMCS theory on the Seifert manifold.
The partition function and the Wilson loops of YMCS theory and CS theory at low energy
can be evaluated exactly by using the localization theorem.
We found that the partition function reduces to the finite dimensional integral
over the eigenvalues of the adjoint scalar field
and the summation over the classical flux configurations.
In the particular cases, the finite dimensional integral further reduces to the
summation over the discrete integer set owing to the Poisson resummation formula.

K\"all\'en's cohomological field theory approach makes clear the relation to
the equivariant cohomology in the localization.
We understand deeper the meanings of the fixed points and constraints
and how to work the localization in the supersymmetric YMCS theory.
The united cohomological approach is easily and  formulated on the 
various topologically distinguished three-dimensional
manifolds possessing the $U(1)$ isometry.

We also derived the supersymmetric indexes of the YMCS theory, which completely coincide
with the expected counting of the brane configurations.
This derivation of the index can be extended to more various CS theories with matters \cite{Suyama:2012kr}
and we can discuss the dynamical supersymmetry breaking for these theories.
It is also interesting to extend these counting to the generalized index like the superconformal index.
We may find the relationships and dualities between the indexes and partition functions in various dimensions
\cite{Kim:2009wb,Terashima:2011qi,Dolan:2011rp,Imamura:2011uw,Gadde:2011ik,Terashima:2011xe,Spiridonov:2011hf,Benini:2011nc}
through the localization of the CS theories on the Seifert manifolds.

We did not specially discuss in the present paper the asymptotic behavior of the partition function 
in the large $N$ limit although it is important to know the M-theoretical nature
of the supersymmetric YMCS theory.
We may discuss the large $N$ limit by using the saddle point approximation
of the matrix model as well as discussed in \cite{Drukker:2010nc,Marino:2011nm,Alday:2012au}.
On the other hand, we find the representation of the partition function
as the summation over the non-colliding discrete integer set.
This fact strongly suggests there exist a suitable form for the large $N$ expansion
at the strong coupling and a free fermion description of the partition function
\cite{Marino:2011eh}. We expect that these understandings of the relationships
may lead to the remarkable Airy function interpretation of the partition function
\cite{Fuji:2011km}.
The large $N$ limit brings us about the relation to the large $N$ reduced matrix models
as discussed in \cite{Ishii:2007sy,Ishiki:2008vf,Asano:2012gt,Honda:2012ni}.
The reduced matrix model deconstructs the planer limit of the continuous field theory
or M-theory in the large $N$ limit.
This might be a good test for the non-perturbative definition of string theory or M-theory.

The cohomological localization can be
extended to five-dimensional contact manifold \cite{Kallen:2012cs}.
We can also discuss the localization of the CS and YM theories
on the topologically distinguished five-dimensional manifolds with the
$U(1)$ isometry, including the case of $S^5$ \cite{Hosomichi:2012ek}.
However, in contrast with the three-dimensional case,
the (physical) supersymmetric gauge theory and topological twisting theory
differ with each other in general
on the higher dimensional manifolds,
since the topological twist changes the spin, as a consequence, the number of the zero modes
from the original theory.
So we need more careful analysis for the higher dimensions, but
the cohomological localization may be still useful for  
some special cases like the circle bundle over the (hyper-)K\"ahler manifolds like K3 surface.
It is interesting to relate the partition functions and indexes of these higher dimensional theories 
with the instanton (BPS soliton) counting or the invariants of the lower dimensional theories.

\section*{Acknowledgements}
We are grateful to 
S.~Hirano,
K.~Hosomichi,
Y.~Imamura,
G.~Ishiki,
S.~Moriyama,
T.~Nishioka,
N.~Sakai,
S.~Shimasaki
and
M.~Yamazaki
for useful discussions and comments.
KO would like to thank the participants in the Summer Institute 2011 in Fuji-Yoshida
and workshops of the JSPS/RFBR collaboration
(``synthesis of integrabilities arising from gauge-string duality'')
for lucid lectures and useful discussions.

\appendix
\section{Topological Twist for Vector Multiplet}
In this appendix we perform the topological twist of the vector multiplet following \cite{Kallen:2011ny}.
The $\mathcal{N}=2$ supersymmetric transformations for
 the vector multiplet on $M$ are
\be
\begin{array}{lcl}
\delta A_{\mu} &=& -\frac{i}{2}(\eb \gamma_\mu \lambda-\tilde{\lambda} \gamma_\mu \e), \\
\delta \sigma &=& \frac{1}{2}(\eb  \lambda-\tilde{\lambda}  \eb), \\
\delta \lambda &=& -\frac{1}{2} \gamma^{\mu \nu} \e F_{\mu \nu}+D \e -i \gamma^{\mu} \e \D_\mu \sigma+ \frac{1}{\ell} \e \sigma, \\
\delta \tilde{\lambda} &=& \frac{1}{2} \gamma^{\mu \nu} \eb F_{\mu \nu} +D \eb - i \gamma^{\mu} \eb \D_\mu \sigma + \frac{1}{\ell} \eb \sigma, \\
\delta D&=&-\frac{i}{2} \eb \gamma^\mu \D_\mu \lambda-\frac{i}{2} \D_\mu \tilde{\lambda} \gamma^\mu \e
+\frac{i}{2} [\eb \lambda +\tilde{\lambda} \e, \sigma ] +\frac{1}{4 \ell} (-\eb \lambda +\tilde{\lambda} \e). 
\end{array}
\label{susyvector}
\ee
We obey the  convection used in \cite{Hama:2010av} except for the Grassmann nature of  $\e$, $\eb$ and  $\delta$.
We take $\e$ and $\eb$ to be the Grassmann even and $\delta$ to be the Grassmann odd.
   
We define the one-form valued twisted fermions $\lambda_\mu$ and $\tilde{\lambda}_\mu$  by
\be
\begin{array}{l}
\lambda =\gamma^\mu \e \lambda_{\mu}, \quad \tilde{\lambda} =\eb \gamma^\mu  \tilde{\lambda}_{\mu}.
\end{array}
\label{twistedgaugino}
\ee
We use gamma matrix identities
\be
\begin{array}{l}
\gamma_\mu \gamma_\nu= g_{\mu \nu}+i \epsilon_{\mu \nu \rho} \gamma^\rho, \\
\gamma_\mu \gamma_{\nu \rho}- \gamma_{\nu \rho} \gamma_\mu=-2 g_{\mu \rho} \gamma_\mu +2 g_{\mu \nu} \gamma_{\rho},  \\
 \{ \gamma_\mu,  \gamma_\nu\}=2 g_{\mu \nu}.
\end{array}
\label{gammaidentities}
\ee
and substitute (\ref{twistedgaugino}) into (\ref{susyvector}), the supersymmetric transformations for the vector multiplet become
\be
\begin{array}{l}
\delta A_{\mu} =-\frac{i}{2}(\eb \gamma_\mu \lambda-\tilde{\lambda} \gamma_\mu \e)=-\frac{i}{2} \Bigl( {\lambda}_\mu-\tilde{\lambda}_\mu  
+i \e_{\mu \nu \rho}\eb \gamma^\rho \e ( {\lambda}^\nu+ \tilde{\lambda}^\nu ) \Bigr), \\
\delta \sigma  =\frac{1}{2}(\eb  \lambda-\tilde{\lambda}  \eb)=\frac{1}{2} \eb \gamma^\mu \e ( {\lambda }_\mu-\tilde{\lambda}_\mu   ), \\
\delta (\eb \lambda -\tilde{\lambda} \e ) =  -i2 \eb \gamma^{\mu} \e \D_\mu \sigma, \\
\delta (\eb \lambda +\tilde{\lambda} \e ) =- \eb \gamma^{\mu \nu} \e F_{\mu \nu} +2D   + \frac{2}{\ell} \sigma, \\
\delta (\eb \gamma_{\mu} \lambda -\tilde{\lambda} \gamma_\mu \e )=-2 \eb \gamma^\nu \e F_{\mu \nu} -2i  \D_\mu \sigma,  \\
\delta D=-\frac{i}{2} \D_\mu ( \eb \gamma^\mu  \lambda+ \tilde{\lambda} \gamma^\mu \e)
+\frac{i}{2} [\eb \lambda +\tilde{\lambda} \e, \sigma ] +\frac{1}{2 \ell} (-\eb \lambda +\tilde{\lambda} \e). 
\end{array}
\ee
Here we also used the normalization condition $\eb \e=1$.
$V^{\mu}=\eb \gamma^\mu \e$ defines a Killing vector along the fiber direction.
If we take constant spinor $\e=(i,0)^t$ and $\eb=(0,i)^t$, the Killing vector becomes
\be
\begin{array}{l}
\eb \gamma^a \e=\delta^{a 3}
\label{killingvector}
\end{array}
\ee
This allow us to take $A_\kappa$, $\lambda_\kappa$ and $\tilde{\lambda}_\kappa$  to be $A_3$ and $\lambda_3$ and $\tilde{\lambda}_3$. 
We now write down the supersymmetric transformation explicitly in each component as
\be
\begin{array}{l}
\delta (A_1+i A_2) =-i(\lambda_1+i \lambda_2 ), \\
\delta (A_1-i A_2) =i(\tilde{\lambda}_1-i \tilde{\lambda}_2 ), \\
\delta A_{\kappa}  =-\frac{i}{2} ( {\lambda}_\kappa-\tilde{\lambda}_\kappa  ), \\
\delta \sigma =\frac{1}{2}  ( {\lambda }_\kappa-\tilde{\lambda}_\kappa   ), \\
\delta ( {\lambda }_\kappa-\tilde{\lambda}_\kappa ) =  -i2  \D_\kappa \sigma, \\
\delta ( {\lambda }_\kappa+\tilde{\lambda}_\kappa ) =-  \e_{\mu \nu \kappa} F^{\mu \nu} +2D   + \frac{2}{\ell} \sigma, \\
\delta (\lambda_1+i \lambda_2 ) =-2  (F_{1 \kappa}+iF_{2 \kappa} ) -2i  \D_\kappa \sigma,    \\
\delta (\tilde{\lambda}_1-i \tilde{\lambda}_2 ) =-2  (F_{1 \kappa}-iF_{2 \kappa} ) -2i  \D_\kappa \sigma,    \\
\delta D=-\frac{i}{2} \D_\kappa ( {\lambda }_\kappa+\tilde{\lambda}_\kappa )
-\frac{i}{2} \Bigl\{ (\D^1-i\D^2)(\lambda_1+i\lambda_2)+(\D^1+i\D^2)(\tilde{\lambda}_1-i\tilde{\lambda}_2) \Bigr\} \\
\qquad \qquad +\frac{i}{2} [{\lambda }_\kappa+\tilde{\lambda}_\kappa , \sigma ] -\frac{1}{2 \ell} ({\lambda }_\kappa-\tilde{\lambda}_\kappa). 
\end{array}
\label{susyvectortwist}
\ee

We now introduce  new fields $\lambda_z $, $\lambda_\zb$, $\eta  $, $\chi_r$ and $Y_r$ by
\be
\begin{array}{lcl}
\lambda_\zb&=&-\frac{i}{2}(\lambda_1+i \lambda_2 ) \\
\lambda_z&=&\frac{i}{2}(\tilde{\lambda}_1-i \tilde{\lambda}_2 ) \\
\eta&=&-\frac{i}{2}  ( {\lambda}_\kappa-\tilde{\lambda}_\kappa  )\\
\chi_r&=&\frac{1}{2} ( {\lambda }_\kappa+\tilde{\lambda}_\kappa )  \\
Y_r&=& -i \e^{\mu \nu \kappa} F_{\mu \nu} +D+\frac{1}{\ell} \sigma
\end{array}
\ee
Rewriting (\ref{susyvectortwist}) by these fields,  we finally obtain the desired scalar BRST transformations (\ref{BRST transformations 1}) :
 \be
\begin{array}{ll}
Q A_{z} =\lambda_z, & Q  \lambda_z=i   F_{z \kappa} -  \D_z \sigma    \\
Q A_{\zb} =\lambda_\zb, & Q  \lambda_\zb =i   F_{\zb \kappa} -  \D_z \sigma   \\
Q A_{\kappa} =\eta, &\\
Q \sigma = i \eta, & Q \eta =  -  \D_\kappa \sigma \\
Q Y_r =-i  \D_{\kappa} \chi_r+ i [\chi_r, \sigma ], &
Q \chi_r =Y_r,
\end{array}
\ee
where we have defined the BRST charge $Q$ from the supercharge $\delta$. 
They are also written in terms of the equivalent form notation:
\be
\begin{array}{ll}
Q A=\lambda, & Q \lambda=-i (-i d_{\A} \sigma + \iota_V F),  \\
Q \sigma = i \iota_V \lambda,  &\\ 
Q Y_r=-i( \iota_V d_{\A} \chi   +[\sigma, \chi_r] ), & Q \chi_r=Y_r.
\end{array}
\ee

\section{Topological Twist for Chiral Multiplet}
Let us next consider topological twist of matter fields.
The supersymmetric transformations for chiral multiplet on $M$ are
\be
\begin{array}{lcl}
\delta X &=& \eb \psi', \\
\delta \psi' &=& -i \gamma^\mu \e \D_\mu X -i \e \sigma X + \frac{\Delta}{\ell}  \e X + \eb F_c, \\
\delta F_c &=& -\e (i \gamma^\mu \D_\mu \psi' - i\sigma \psi' - i \lambda X)-\frac{1}{2\ell} (2 \Delta-1) \e \psi'. 
\end{array}
\label{susychiral}
\ee
We define the topologically
twisted one-form fermion $\psi_{\mu}$ similar manner as vector multiplet by
\be
\begin{array}{l}
 \psi'\equiv \gamma^\mu \e {\psi}_\mu. 
\end{array}
\label{twistedsquark}
\ee
Substituting (\ref{twistedsquark}) into (\ref{susychiral}) and using (\ref{gammaidentities}), (\ref{killingvector}) and $\eb \eb=\e \e=0$,
we obtain
\be
\begin{array}{l}
\delta  X =   {\psi}_\kappa, \\
\delta   \psi_\kappa =-i  \D_\kappa X -i  \sigma X + \frac{\Delta}{\ell}  X,  \\
\delta \bigl(  \psi^\mu +i \e^{\mu \nu \kappa } \psi_\nu \bigr)
 =-i \bigl( \D^\mu X +i \e^{\mu \nu \kappa } \D_\nu X  \bigr) -i \eb \gamma^\mu \e \sigma X 
+ \frac{\Delta }{\ell} \eb \gamma^\mu \e X + \eb \gamma^\mu \eb F_c, \\
\delta F_c  = \e^{\mu \nu \rho} \e \gamma_\rho \e \D_\mu \psi_\nu
+ i\sigma \e \gamma^\mu \e \psi_\mu + i \e \gamma^\mu \e \lambda_\mu X
-\frac{\Delta}{\ell}    \e \gamma^\mu \e   \psi_\mu. 
\end{array}
\ee
We  define  new fields $\psi$ belongs to  zero-form, $\chi_{\zb}$ and $Y_{\zb}$ belong to $(0,1)$-form by
\be
\begin{array}{lcl}
\psi &=& \psi_\kappa, \\
\chi_{\zb} & = & \psi_1+i \psi_2, \\
Y_{\zb} & = & -(F_c+i(\D_1+i \D_2) X),
\end{array}
\ee
and use the relations 
\be
\begin{array}{lll}
\e  \gamma^1 \e=-1, &
\e \gamma^2 \e=-i, &
\e \gamma^3 \e=0, \\
\eb \gamma^1 \eb=1, &
\eb \gamma^2 \eb=-i, &
\eb \gamma^3 \eb=0,
\end{array}
\ee
we obtain the scalar BRST transformations in the new variables:
\be
\begin{array}{ll}
Q  X =  \psi, & Q \psi=-i \D_{\kappa} X -i \sigma X +\frac{\Delta}{l} X, \\
Q Y_{\zb}=-i ( \D_{\kappa} +\sigma ) \chi_{\zb} +\frac{\Delta}{l} \chi_{\zb},  &  Q \chi_{\zb}=Y_{\zb}.   
\end{array}
\ee
As explained in the article, if we include the inhomogeneous term
proportional to $\Delta/\ell$ 
into the eigenvalue of $\del_\kappa$ ($\cL_V$)
by imposing the twisted boundary condition,
the BRST transformations reduce to
the compatible form with the vector multiplet
\be
\begin{array}{ll}
Q  X =  \psi, & Q \psi=-i \D_{\kappa} X -i \sigma X, \\
Q Y_{\zb}=-i ( \D_{\kappa} +\sigma ) \chi_{\zb},  &  Q \chi_{\zb}=Y_{\zb}.   
\end{array}
\ee


\end{document}